\documentstyle[epsfig]{mn}
\def\etal{{\it et al}}
\def\deg{^{\circ}}
\def\parhang{\noindent\hangindent=0.4 true in \hangafter=1}
\newcount\notenumber
\def\clearnotenumber{\notenumber=0}
\def\note{\advance\notenumber by1 \footnote{$^{\the\notenumber}$}}
\clearnotenumber
\begin{document}
\title[PSR B0943+10's ``Drifting''-Subpulses I.] 
  {Topology and Polarisation of Subbeams Associated With \\
   Pulsar 0943+10's ``Drifting''-Subpulse Emission:  \\
   I. Analysis of Arecibo 430- and 111-MHz Observations}

\author[Deshpande \& Rankin] 
{Avinash A. Deshpande$^{1}$~~\&~ Joanna M. Rankin$^{2}$ \\ 
$^1$ Raman Research Institute, Bangalore 560 080 INDIA : desh@rri.ernet.in \\
$^2$ Physics Department, University of Vermont, Burlington, VT 05405 USA : rankin@physics.uvm.edu }
%\begin{document}
\date{}
\maketitle

\begin{abstract}
The ``drifting'' subpulses exhibited by 
some radio pulsars have fascinated both 
observers and theorists for 30 years, and 
have been widely regarded as one of the 
most critical and potentially insightful 
aspects of their emission.  Moreover, 
Ruderman \& Sutherland (1975), in their 
classic model, suggested that such regular 
modulation was produced by a system of 
subbeams, rotating around the magnetic 
axis under the action of $\bf{E}$$\times$$\bf{B}$ 
drift.  Such ``drift'' sequences have 
thus been thoroughly studied in a number 
of pulsars, but it has proven difficult 
to verify the rotating-subbeam hypothesis, 
and thus to establish an illuminating 
connection between the phenomenon and 
the actual physics of the emission.  

Here, we report on detailed studies of 
pulsar B0943+10, whose nearly coherent 
sequences of ``drifting'' subpulses 
have permitted us to identify their 
origin as a system of subbeams that 
appear to circulate around the star's magnetic 
axis.  We introduce several new 
techniques of analysis, and we find 
that both the primary and secondary 
features in the star's fluctuation 
spectra are aliases of their actual 
values.  We have also developed a 
method of tracing the underlying pattern 
responsible for the observed sequences, 
using a ``cartographic'' transform and 
its inverse, permitting us to study 
the characteristics of the polar-cap 
emission ``map'' and to confirm that 
such a ``map'' in turn represents 
the observed sequence.  We apply these 
techniques to the study of three 
different Arecibo observations: a 1992 
430-MHz sequence which includes a 
transition from the star's highly 
organized ``B'' profile mode to its 
disorganized ``Q'' mode; a 1972 
430-MHz ``B''-mode sequence; and a 
1990 111-MHz ``B''-mode sequence.  

The ``B''-mode sequences are consistent 
in revealing that the emission pattern 
consists of 20 subbeams, which rotate 
around the magnetic axis in about 37 
periods or 41 seconds.  Even in the 
``Q'' mode sequence, we find evidence 
of a compatible circulation time.   The 
similarity of the subbeam patterns at 
different radio frequencies strongly 
suggests that the radiation is produced 
within a set of columns, which extend 
from close to the stellar surface up 
though the emission region and reflect 
some manner of a ``seeding''phenomenon 
at their base.  The  
subbeam emission is then tied neither 
to the stellar surface nor to the 
field.  While the origin of the 
``memory'' responsible for the 
stability of the pattern over several 
circulation times is unknown, the 
hollow-conal form of the average 
pattern is almost certainly the origin 
of the conal beam forms observed in 
most pulsars.

\end{abstract}

\begin{keywords}
MHD --- plasmas --- pulsars: general, individual (B0943+10) 
--- radiation mechanism: nonthermal --- polarisation
\end{keywords}

\eject
\eject

\section*{I. Introduction}

We wish to reopen several questions which have 
lain dormant for most of 20 years:  What is responsible 
for the spectacular sequences of ``drifting'' 
subpulses observed in many pulsars with conal 
single ({\bf S}$_{\rm d}$) profiles?  What is the 
topology and polarisation of the individual 
beaming elements which correspond to these 
subpulses---and how do they accrue to form the 
average profile?  How can we understand a range 
of other observed effects---profile and polarisation 
modes, multiple $P_3$ values, ``absorption'' and 
subpulse ``memory''---in terms of fundamental emission 
processes and propagation effects within the pulsar 
magnetosphere?  

From the initial identification of ``drifting 
subpulses'' by Drake \& Craft (1968), many suspected 
the phenomenon to be an unusually clear manifestation of 
the physical processes behind pulsar emission.  
Fluctuation-spectral analysis (taken over from 
scintillation studies) found its place almost 
immediately in assessing pulse sequences for 
periodicity (Lovelace \& Craft 1968), and 
strenuous efforts were made by several groups 
to study the subpulse modulation in the then 
known stars ({\it i.e.}, Taylor \etal\ 1969; Lang 
1969; Cole 1970; Sutton \etal\ 1970; Slee \& Mulhall 
1970).  Backer's (1970{\it a-c},1971,1973) thesis 
research carried these early efforts to understand 
drifting subpulses to a new level; he systematized 
the then existing studies, clarified how sightline 
geometry could produce distinct modulation 
characteristics, developed the crucial technique of 
applying fluctuation-spectral analysis to each narrow 
longitude interval within the emission window (and 
of following the varying phase of a feature with 
longitude), and introduced the terms within which 
the ``drift'' phenomenon is now almost universally 
understood.  

Of course, the very existence of individual subpulses 
raises unresolved questions about how these distinct 
elements of emission are produced---but stars with 
periodic subpulse modulation (drifting in single 
stars with single profiles and longitude-stationary 
modulation in those with double ones) have suggested, 
from near the time of their discovery, a system of 
regularly spaced subbeams which rotate progressively 
around the magnetic axis of the star (Ruderman 1972).  
Several efforts have been made to delineate the beam 
topology and polarisation characteristics of these 
subbeams, but none have thus far been very successful.  
Pulsar B0809+74 has attracted special attention because 
its subpulse modulation is so remarkably regular (Sutton 
\etal\ 1970; Taylor \& Huguenin 1971)---giving a strong, 
high-``Q'' feature at some 0.090 cycles/period in its 
fluctuation spectrum.  However, the most creative 
efforts at measuring the polarisation of its 
individual-pulse sequences have resulted in only 
marginal signal strength relative to the noise level 
(S/N), so that these characteristics are now known 
only in the average of several drift sequences (Taylor 
\etal\ 1971).  

Pulsar B0943+10 is another star in 0809+74's class.  
Its fluctuation spectra also exhibit narrow 
features---this time at about 0.47 cycles/period---but 
its unusually steep spectrum makes the star difficult 
to observe on a single-pulse basis at frequencies 
above 300 MHz.  Its pulse-to-pulse modulation is then 
nearly odd-even, with alternate pulses appearing to 
progress slowly from the trailing to leading edge of 
the window over 15 or so periods.  No entirely 
consistent picture has emerged from the existing 
studies of its drift phenomenon.  Taylor \& Huguenin 
(1971) first associated the drift pattern with another, 
weaker fluctuation feature at 0.065 cycles/period and 
attributed the stronger feature mentioned above to 
the nearly ``alternate pulse modulation''.  Backer 
(1973) noted the general possibility that such a 
primary feature was an alias of the actual fluctuation 
frequency, but found no means of distinguishing between 
the possibilities.  Backer \etal\ (1975) then showed 
that the two features have an exact harmonicity if 
the secondary one is the aliased second harmonic of 
the primary one, but then went on to discuss the phase 
function of the primary feature without considering 
whether it also might be aliased.  Finally, Sieber \& 
Oster (1975) clearly discuss some of the aliasing 
possibilities, attempt to distinguish between them, 
but (in our view) come to the wrong conclusion---and 
thus find no cause to mention any harmonicity between 
the features.  We note that these various studies 
measured significantly different frequencies for the 
primary feature, which are summarized in Table I.  

\pagestyle{myheadings}
\markboth{Deshpande \& Rankin}{Pulsar 0943+10's ``Drifting''-Subpulse Emission}

PSR 0943+10 exhibits two distinct profile ``modes'' 
at 102 MHz, designated ``B'' and ``Q'' by their 
discoverers, Suleymanova \& Izvekova (1984).  In 
an earlier paper (Suleymanova \etal\ 1998), the 
properties of these modal sequences are explored at 
430 MHz for the first time using an exceptionally 
strong, 18-minute polarimetric observation.  The 
contrasting characteristics of the two modes are 
striking.  The B (for ``bright'') mode entails a 
steady, highly organized pattern of subpulse 
emission---the drifting subpulses discussed 
above---which is confined to the early part of the 
emission window; whereas the weaker ``Q'' (for 
``quiescent'') mode produces disorganized, but 
occasionally much more intense, subpulses throughout 
the entire window.  The primary and secondary 
polarisation modes (PPM and SPM, respectively) are 
emitted in both B- and Q-mode sequences, the former 
well dominated by the PPM and the latter only slightly 
dominated by the SPM, respectively---resulting in 
different levels of aggregate linear depolarisation.  

These circumstances raise many questions about how 
the pulsar's average, polarised modal profiles are 
constituted of their respective pulse sequences, 
which were only partially addressed in the earlier   
paper.  Furthermore, these subpulses directly reflect 
the electrodynamic processes responsible for pulsar 
emission---that is, (according to the received 
``cartoon'') primary particle acceleration associated 
with local ``sparking'' regions near the polar cap, 
which in turn generate a secondary plasma ``bunches'' 
that radiate coherent radio emission.  We will attempt 
to relate our observations and analysis to the 
general picture elaborated by Ruderman \& Sutherland 
(1975; hereafter R\&S).  

Finally, something is known about 0943+10's overall 
emission geometry.  Estimates of its magnetic 
latitude angle $\alpha$ and impact angle $\beta$ 
were made by one of us (Rankin 1993{\it a,b}) on 
the basis of the best information then available.  
It was clear even from these initial values that our 
sightline barely grazes the pulsar's emission cone, 
reconfirming its classification as having a conal 
single ({\bf S}$_{\rm d}$) profile geometry.   Moreover, 
at 400 MHz and above, the pulsar's decreasing conal 
beam radius is apparently responsible for the pulsar's 
extremely steep spectral decline (Comella 1971).  
Only recently have weak detections been reported at 
frequencies above 600 MHz: Deshpande \etal\ (1999) 
at 840 MHz and Weisberg \etal\ (1999) at 21 cms.  At 
and below 100 MHz the pulsar is one of the brightest 
in the sky; indeed, to our knowledge it is the first 
and still the only pulsar discovered at a frequency 
below 300 MHz (Vitkevitch \etal\ 1969), and it is 
unique in exhibiting no spectral turnover down 
to some 25 MHz (Izvekova \etal\ 1981; Deshpande 
\& Radhakrishnan 1992).\footnote{Malofeev (1999) now 
has evidence suggesting that the pulsar's spectrum 
begins to turn over between 34 and 25 MHz.} 
\S II discusses our observations, and \S\S III--IV  
the star's modulation features and aliasing.   \S VI 
assesses the star's emission geometry.  \S VII 
discusses our subbeam imaging technique and applies 
it to the 1992 ``B''-mode sequence.  The ``Q''-mode 
sequence is discussed in \S VIII.  \S IX discusses 
the polarization-mode structure of the ``B''-mode 
sequence.  
\S\S X--XI discuss the older (1974) 430-MHz sequence 
and a more recent 111.5-MHz one.  Finally, \S XII 
briefly summarizes the arguments and 
conclusions.\footnote {Some early results of 
this paper appear in Deshpande \& Rankin (1999), 
Deshpande (1999), and Rankin \& Deshpande (1999).}  

%<    I. Introduction (in draft)
%<   II. Observations (partly drafted)
%<  III. Oct92 Primary Features and Aliasing (drafted)
%<   IV. Spacing of the Drifting Subbeams (drafted)
%<    V. Oct92 Amplitude Modulation ===> 20 Sparks (drafted)
%<   VI. Overall Emission Geometry (drafted)
%<  VII. Oct92 Imaging the Polar Cap (drafted)
%< VIII. Oct92 Q mode sequence 
%<   IX. Polarisation of the Subbeam System
%<    X. 1972 430-MHz Sequence
%<   XI. 111.5-MHz Sequence
%<  XII. Facing up to Theory
%< XIII. Modes, Sparks, Drift \& ``Absorption''---Building a
%<          Cartoon of 0943+10's Emission
%<  XIV. Discussion: Implications of Other Pulsars/Observations
%<   XV. Discussion: Summary and Conclusions

\section*{II. Observations}

The single-pulse observations used in our analysis 
below come from two programs carried out at the Arecibo 
Observatory over a long period of time.  The newer 
430-MHz observation was made at the Arecibo 
Observatory on the 19th October 1992 and is identical
to that considered in the companion paper (Suleymanova
\etal\ 1998).  Use of 10-MHz bandwidth, across which
32 channels were synthesized by the 40-MHz Correlator, 
and a 1006 ${\mu}$s dump time reduced dispersion delay 
across the bandpass to negligible levels.  The 
resolution was then essentially the dump time of 
$0.328\deg$.  The observational procedures will be 
described in a forthcoming paper (Rankin, Rathnasree 
\& Xilouris 2000).

The older 430-MHz observation was carried out on the 
2nd January 1972 with a single-channel polarimeter of 
2-MHz bandwidth and 1-ms integration time, giving 
a nominal time resolution of about $1.1\deg$ longitude. 
The polarimetry scheme is described in Rankin \etal\ (1975).  

The 111.5-MHz observation was made 
on the 17th January 1990 in an earlier 
phase of the 40-MHz-Correlator-based polarimetry 
program.  Use of 2.5-MHz bandwidth and retention of 
64 lags (channels) along with a 1406 ${\mu}$s dump 
time resulted in an effective resolution of about 
$1.265\deg$.  No continuum source observation is 
available to calibrate Stokes parameter $V$ for this 
observation, so the baseline levels, which were 
dominated by galactic background noise, were thus 
used to calibrate the relative gains of the channels.

\section*{III. Primary Fluctuation Features}

We first compute nominal 256-point fluctuation 
spectra for 0943+10's ``B'' mode, using the 816-pulse 
sequence at 430 MHz discussed in Suleymanova \etal, 
and the result is shown in Figure~\ref{fig:fig1}.  The slow 
(about 30\%) decrease of intensity over this interval 
(which they noted) was flattened before computing 
the spectra to avoid possible smearing of the otherwise 
high-Q features.  As expected, these longitude-resolved 
fluctuation spectra (hereafter, ``LRF spectra'') show 
a strong feature near 0.46 cycles/rotation period 
(hereafter, c/$P_1$), which most fully modulates the 
power on the leading edge of the profile, but whose 
effect can be  discerned throughout the ``B''-mode profile.  
The remarkable strength and narrowness of this feature 
is evident in the integral spectrum, from which its 
frequency can be accurately determined as 
$0.465\pm0.001$ c/$P_1$.  This value agrees well with 
that determined previously by Backer \etal\ (1975), 
but is incompatible with those of Taylor \& Huguenin 
(1971) and Sieber \& Oster (1975) (which, paradoxically, 
are more accurately determined and consistent with 
each other); see Table I.  The half width of the 
integral feature is less than 0.002 c/$P_1$, making its 
``Q'' more than 200!  No known pulsar, including 
0809+74, has such a remarkably stable modulation 
feature.

\begin{figure*}
\begin{center}
Table I.  Modulation Feature Frequencies
\begin{tabular}{lcccc}\hline\hline
Source & $f_p$(c/$P_1$) & $f_p$ alias & $f_s$(c/$P_1$) & harmonic? \\
\hline
Taylor \&Huguenin & $0.477\pm0.003$ & not cons'd  & $0.065\pm0.005$ & no \\
Backer \etal\ & $0.461\pm0.010$ & not cons'd & $0.078\pm0.020$ & exact \\
Sieber \& Oster & $0.473\pm0.002$ & $0.527\pm0.002$  & $0.07\pm0.01$ & not cons'd  \\
1992/430 MHz & $0.4645\pm0.0003$ & $0.5355\pm0.0003$ & $0.0710\pm0.0007$ & exact  \\
1972/430 MHz & $0.4591\pm0.0009$ & $0.5409\pm0.0009$ & not determined & ---  \\
1990/111 MHz & $0.4688\pm0.0005$ & $0.5312\pm0.0005$ & not determined & ---  \\
\hline
\end{tabular}
\end{center}
\end{figure*}

\begin{figure}
\epsfig{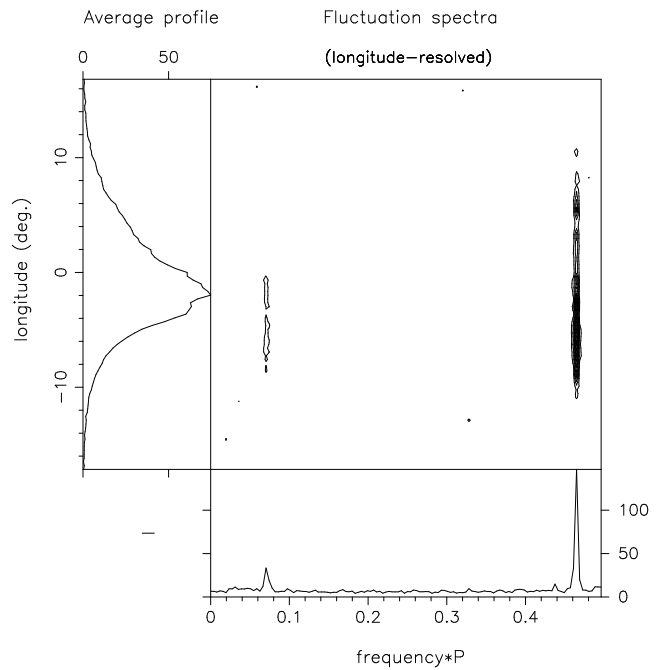}
%\parhang {\bf Figure 1.}---
\caption{Longitude-resolved
fluctuation spectrum for pulsar 0943+10 at
430 MHz.  A 256-point fft was used and
averaged over the first 816 pulses of the
19 October 1992 observation.  The body of
the figure gives the amplitude of the
features, with contours at intervals of
0.012 mJy$^2$, which reach a maximum value
of 0.11 mJy$^2$.  The average profile
(Stokes parameter $I$) is plotted in the
left-hand panel, and the integral spectrum
is given at the bottom of the figure.  The
pulse-sequence amplitude was adjusted before
analysis to remove the slow secular decrease
[which is intrinsic to the pulsar, see
Suleymanova \etal\ (1998)], so that any
small $1/f$ ``tail'' in the spectrum has
been suppressed.  Note that four distinct
features are clearly visible in the integral
spectrum, the two major features as well as
a pair of weak features on either side of
the principal one (see text).
\label{fig:fig1}}
\end{figure}

\begin{figure}
\epsfig{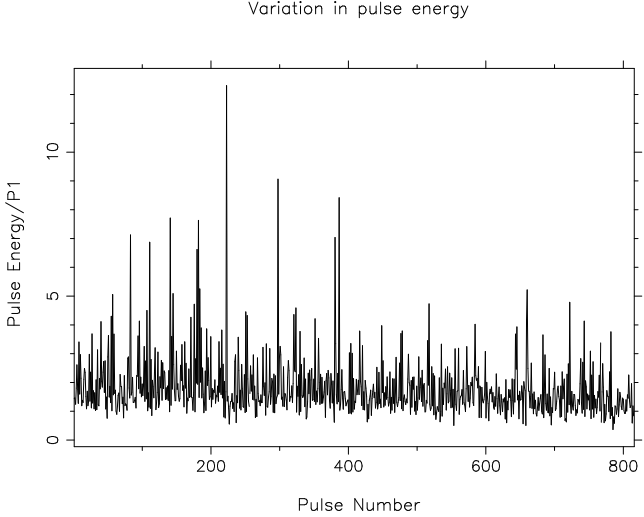}
%\parhang {\bf Figure 2.}---
\caption{Pulse energy as a function of pulse 
number of the 816 pulses comprising the 
``B''-mode sequence.  Notice the slight secular 
decrease in intensity, its periodic tendencies, 
and, most of all, notice that there is not a 
single ``null'' pulse!  
\label{fig:fig2}}
\end{figure}

Nonetheless, the principal feature in Fig.~\ref{fig:fig1} 
is not resolved with a Fourier transform of length 
256.  Longer FFTs yield a progressively narrower 
feature, and even for a transform of length 816, the 
principle feature is only partly resolved.  On this 
basis, we obtain an $f_3$ value of $0.4645\pm0.0003$ 
c/$P_1$, which implies a ``Q'' greater than 500.  
Subpulse modulation with such remarkable stability is 
unprecedented.  Indeed, we cannot yet say whether this 
behaviour is typical or unusual even for 0943+10, but 
during this 14.9-minute ``B''-mode sequence, its 
subpulse modulation has a bell-like quality---and we 
see in Figure~\ref{fig:fig2} that the sequence is 
undisturbed by even a single null pulse.  

A second feature is seen near 0.07 c/$P_1$, and it 
modulates the emission early in the B-mode profile just 
as did the primary feature above.  We calculate the 
frequency of this smaller feature as $0.0710\pm0.0007$ c/$P_1$.  
We note also that this feature's width is larger, about 
twice that of the principal one.  An FFT of 
length 512 yields the slightly improved (mean) value 
of $0.0697\pm0.0005$ c/$P_1$---and clearly resolves 
the width of the feature.  

The possible relationship between the two spectral 
features warrants detailed discussion.  Of course, 
it is possible to speculate, as did Sieber \& Oster, 
that one or both of the features are aliases of 
fluctuations with frequencies greater than 0.5 
cycles/period.  Indeed, we can then confirm that if 
the principal feature is the first-order alias 
($1 - f_3 = 0.4645\pm0.0003$ c/$P_1$) of a fluctuation 
whose actual frequency $f_3$ is $0.5355\pm0.0003$ c/$P_1$, 
then its  second harmonic $f_3^\prime = 2 f_3$ would 
fall at $1.0710\pm0.0006$ c/$P_1$; and this fluctuation 
would then have a second-order alias at ($f_3^\prime 
[alias] = f_3^\prime -1$) at $0.0710\pm0.0007$ c/$P_1$.  
If, however, the secondary feature appears as a 
first-order alias, the two might still have an harmonic 
relationship, because $f_3^\prime$ would be 
$0.9290\pm0.0006$ c/$P_1$, which would 
then have a first-order alias 
($1 - f_3^\prime = $) $0.0710\pm0.0008$ c/$P_1$.  
The measured frequencies accurately support the 
premise of an harmonic relationship as long as one or 
both are aliased as above. Or, as Backer \etal\ concluded, 
it was ``possible to avoid the conclusion of an harmonic 
relationship only by assuming that the relationship is 
fortuitous''.  In fact, even higher order aliases in suitable 
combinations would be consistent with the premise of 
an harmonic relationship.  Therefore, we can be certain 
about the harmonicity of the two features only if we 
establish their alias orders or---equivalently---the 
``true`' frequencies of the features, a desired 
information in any case.  

In order to do so, we need to sample the fluctuations 
faster than just once a period.  This may be possible 
by appealing to the fact that the fluctuations are 
sampled also within the finite width of the pulse. 
In other words, the fluctuation spectra at different 
longitudes can be combined appropriately based on their 
longitude separations to ``unfold'' the contributions 
that are otherwise aliased within the 0--0.5 c/$P_1$ band.
In practice, we can achieve this simply by Fourier 
transforming the entire time sequence, which can be 
reconstructed using the available pulse sequence and 
filling the unsampled (off-pulse) regions with zeros. 
Such (power) spectra can be computed for pulse sequences
in suitable blocks and 
then averaged.  Care must be taken to ensure that the 
sampled region is wide enough to include the entire 
pulse-emission window ({\it i.e.} untruncated) and that
``baseline'' is removed correctly, so that the spectral 
features corresponding to the pulsar signal are not 
distorted. This effective windowing of the time sequence, 
where only the off-pulse contributions are forced to 
zero, leads to a convolution of the spectrum of the 
raw (unwindowed) time sequence with the spectrum of the 
periodic windowing function.  However, it is easy to 
show that this convolution does not modify the spectral 
contribution of the pulsar signal, smoothing only the
noise and hence, in fact, enhancing the signal-to-noise 
ratio (hereafter, ``S/N'') of the spectral features 
of interest.

\begin{figure}
\epsfig{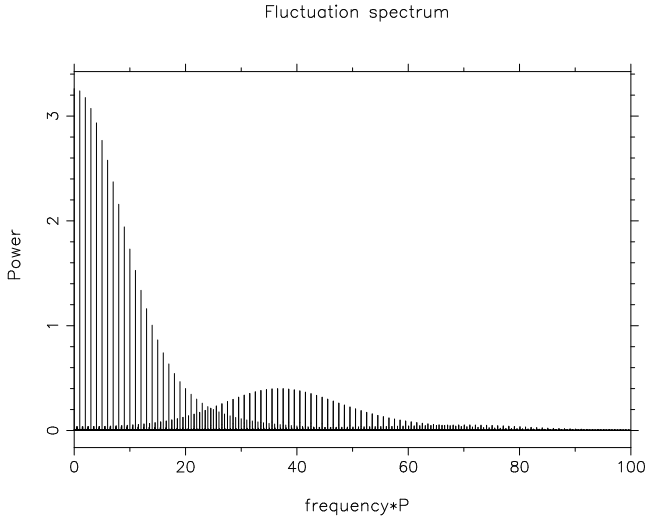}
%\parhang {\bf Figure 3.}---
\caption{An unfolded fluctuation spectrum, obtained 
by continuously sampling the same sequence as in 
Fig.~\ref{fig:fig1}.  Three successive, overlapping, 
256-pulse Fourier transforms (each of length about 
$2.8\times10^5$ points, when the unsampled 
intrapulse region was interpolated with zeroes) were 
averaged together.  There are three main contributions 
to this diagram: a) Harmonics at integral multiples of 
the fundamental $1/P_1$ pulsar-rotation frequency, 
which are related to the Fourier components of the 
``average'' profile.  These are strongest at low harmonic 
numbers and have declined to a relative amplitude of 
$1/e$ by about number 15.  b) Components falling at
about half-integral frequencies [$(N+0.535)/P_1$], 
which peak at about harmonic 35, and c) a set of 
components at frequencies near (but not exactly at) 
harmonics of $1/P_1$ [$(N+0.071)/P_1$], which
peak near number 70.  The latter two sets of
components represent the harmonics of the
highly periodic fluctuations associated with
the pulsar's subpulse modulation.
\label{fig:fig3}}
\end{figure}

The results of such a computation are given in 
Figure~\ref{fig:fig3}, where only one half of the 
symmetric power spectrum is plotted as a function 
of frequency, normalized to the pulsar rotation 
frequency $P_1$.  There is much to see in this 
diagram.  First, note the amplitude of the frequency 
components at integral multiples of the pulsar's 
rotation frequency.  These are the Fourier 
components of its ``average'' profile, which 
include contributions from its ``base'' profile---or 
what is unchanging from pulse to pulse---as well 
as from its fluctuations seen on an average.  
The harmonics decline monotonically to a $1/e$ 
amplitude by about harmonic number 15; note that 
this corresponds approximately to the longitude 
width of 0943+10's profile.  Now notice the 
spectral components at about half-integral 
frequencies (actually $N+0.535$ c/$P_1$), which 
grow steadily and peak around harmonic number 
35---a scale which corresponds approximately to 
the width of the individual drifting subpulses, 
and more directly to $P_2$, the longitude 
separation between adjacent subpulses.  Finally, 
notice the near-integral frequency components 
(actually $N+0.071$ c/$P_1$) which begin to be 
distinguishable from the precisely integral 
components at harmonic numbers above 50 and 
which peak at about harmonic 70; these represent 
the Fourier components of the secondary feature.  
The relationship ({\it i.e.} the relative 
amplitude and phase) between these sets of 
modulation components define the average ``shape'' 
of the ``drifting'' subpulses, while that among 
the components within a set define the ``envelope'' 
followed by the subpulse as it ``drifts'' in 
longitude.

\begin{figure}
\epsfig{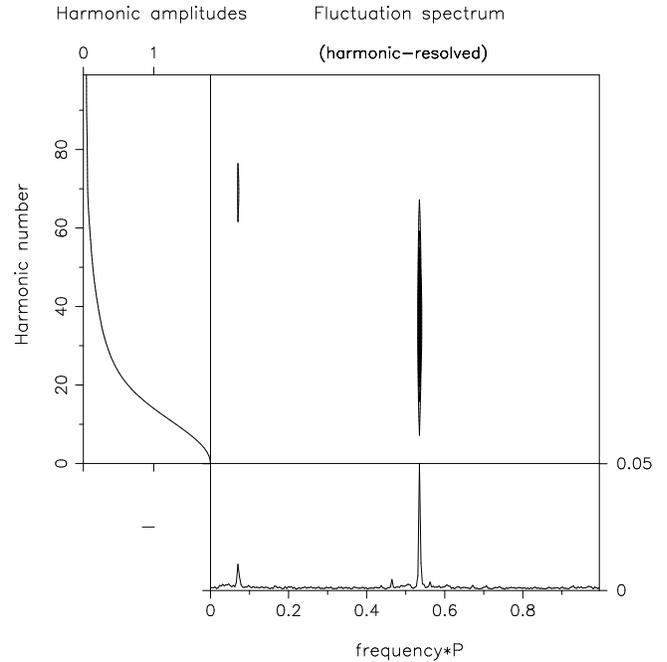}
%\parhang {\bf Figure 4.}---
\caption{An unfolded Fourier 
spectrum as in Fig.~\ref{fig:fig3}.  The 
amplitude of the frequency components at 
the fundamental rotation frequency $1/P_1$ 
and its harmonics are essentially the 
Fourier components of the ``average'' 
profile; and these are plotted in the 
left-hand panel of the diagram.  The body 
of the figure gives the amplitude of all 
the other frequency components in the 
spectrum up to $100/P_1$ as a contour plot 
(with contours at amplitude intervals of 
0.040 mJy$^2$ which reach a maximum value 
of 0.3530 mJy$^2$), and the bottom panel 
shows the sum of these frequency components, 
collapsed onto a $1/P_1$ interval.  The 
principal feature now falls at its true 
(unaliased; see text) frequency of about 
0.535 cycles/period with its first harmonic 
at about 1.071 cycles/period.  We also see 
two ``sidelobes'' associated with the 
principal feature as well as a slight 
``leakthrough'' at its alias frequency 
(from the unplotted negative-frequency 
portion of the spectrum). 
\label{fig:fig4}}
\end{figure}

Just to make things clearer, we present the 
fluctuation spectrum of Fig.~\ref{fig:fig3} 
in a different manner by stacking successive 
sections of $1/P_1$ width one above the other 
as shown in  Figure~\ref{fig:fig4}.  We see 
immediately that the principal feature now 
falls at a frequency of about 0.535 c/$P_1$.  
The power at a frequency of $1/P_1$ and its 
harmonics---which comprise the ``average'' 
profile of the pulsar---are shown in the 
right-hand panel.  The fluctuation power at 
other frequencies, as a function of $1/P_1$-frequency 
interval, is shown in the body of the figure, 
and the integral of this fluctuation power, 
summed into a single $1/P_1$ band, is given 
in the bottom panel. This spectral representation, 
which we will refer to as an ``harmonic-resolved 
fluctuation spectrum'' (hereafter ``HRF 
spectrum''), can also distinguish between the 
two types of modulation encountered, namely 
amplitude- and phase-modulation. For example, 
the summed spectrum (as in the bottom panel) 
would exhibit nearly symmetrical modulation 
sidebands in the case of amplitude modulation, 
while a clear asymmetry will be apparent for 
phase modulation, as is presently the 
case.\footnote{Were, for instance, our 
sightline to make a central traverse across 
0943+10's polar cap, rather than the highly 
tangential one it does, we would expect to see 
the same primary modulation feature.  However, 
it would appear as a longitude-stationary 
modulation---that is, as an amplitude modulation 
without ``drift''---and its fluctuation power 
would thus be divided between positive- and 
negative-frequency features corresponding 
to ``drift'' to the right and to the left.  
Such pairs of identical features at positive 
and negative frequencies reliably identify 
the presence of amplitude modulation.}  So, 
although the two modulation types are in 
general mixed, the asymmetry between the 
0.535 and 0.465 c/$P_1$ features in 
Fig.~\ref{fig:fig4} is so great that we can 
regard them as the signature of an almost 
``pure'' phase modulation.  

Fig.~\ref{fig:fig3} presents {\it exactly} the 
same information as in Fig.~\ref{fig:fig4}, but 
in a slightly more direct way.  Note that 
nothing has been suppressed here.  The two 
figures give the amplitude of all the Fourier 
components of the PS, and its power falls 
predominantly in one of the three sets of 
components discussed above.  Remarkably, the 
aggregate power in the non-integral components 
is comparable to that in the integral ones.  
The Fourier components of the noise are not 
discernible in either Figs.~\ref{fig:fig3} or 
\ref{fig:fig4}, implying that its power (as 
expected) is distributed uniformly over 
fluctuation frequency and is thus completely 
negligible in any given frequency 
interval\footnote{Figs.~\ref{fig:fig3} and 
\ref{fig:fig4} also exhibit the exceptional 
quality of the Arecibo 40-MHz-Correlator-based  
PS observations.  Were the noise not virtually 
``white'' and free of periodic contributions, 
their effects would stand out sharply in these 
diagrams.}  Again, Fig.~\ref{fig:fig3} clearly 
demonstrates the ``bell like'' quality of this 
unprecedented ``B-mode'' sequence from pulsar 
0943+10.  

We can now be certain that the principal feature 
in Fig.~\ref{fig:fig1} is the alias of a 
fluctuation whose actual frequency is greater 
than 0.5 c/$P_1$ and that the secondary feature 
is its second harmonic.  From Figs.~\ref{fig:fig3} 
\& ~\ref{fig:fig4}, it is clear that these 
features, seen in the LRF spectra, are indeed 
higher-order aliases of the two sets of 
frequency components whose distances from zero 
frequency are harmonically related---that is, with 
maximum intensities at about harmonic 35 and 70, 
respectively.  Using a 512-point fft and 
interpolating the peak positions, we determine 
that the two responses have frequencies $f_3$ of 
$0.5352\pm0.0006$ c/$P_1$ and $f_3^\prime$ of 
$1+0.0710\pm0.0006$ c/$P_1$---accurately demonstrating 
their harmonic relationship.  Further, we see 
that the secondary feature has a well resolved 
width of about 0.0006 cycles/period, more than 
twice that of the primary periodicity.  Finally, 
given the strength and narrowness of the primary 
feature's second harmonic, it is not ridiculous 
to explore the possibility of detecting its 
third harmonic, and a quick calculation 
($f_3^{\prime\prime} [alias] = 3 f_3 - 1$) 
confirms that it would fall at some 
$0.6065\pm0.0012$ c/$P_1$, where indeed, a minor 
feature can be seen in Fig.~\ref{fig:fig4}.  

\begin{figure}
\epsfig{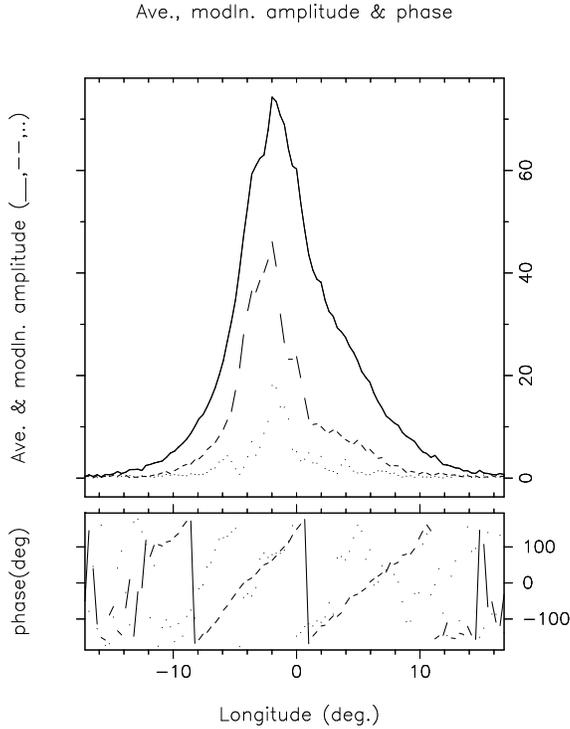}
%\parhang {\bf Figure 5.}---
\caption{Fluctuation-phase and amplitude for 
the primary (0.535 c/$P_1$---dashed curve) 
and secondary (1.070 c/$P_1$---dotted curve) 
features in Figs.~\ref{fig:fig3}--\ref{fig:fig4}. 
Fluctuation power is plotted as a function of 
longitude within the average profile in the top 
panels, whereas the fluctuation phase is given 
in the lower panels. 
\label{fig:fig5}}
\end{figure}

One other line of evidence can be brought to 
bear the aliasing of the observed features---that 
is, the fluctuation phase rate as a function 
of longitude associated with each of them.   
Their respective phase rates can be used to 
assess the harmonicity of the features, as 
well as to determine the drift-band spacing 
($P_2$) in longitude.  Figure~\ref{fig:fig5} 
gives average profiles showing both the 
amplitude and the phase of the power associated 
with the primary and secondary features.  The 
maximum rate in the former is just less than 
$36\deg/\deg$ near the longitude origin, which 
is compatible with a longitude interval between 
subpulses $P_2$ of just over $10\deg$---say, 
$10.5\deg$.  Note that the phase rate is less 
steep under the ``wings'' of the profile, as 
might be expected if the subpulses move along 
a curved path.  The phase rate of the secondary 
feature is much faster and seems to reach some 
$70\deg/\deg$ near the peak of the profile, 
where the fluctuation power also has a maximum.  
We have not reproduced here a similar figure 
for the ostensible third harmonic, which 
exhibits a phase rate nearly thrice that of 
the primary feature.  The harmonicity of these 
phase rates argues strongly that the 0.071 and 
0.607 c/$P_1$ features are indeed, the aliases 
of the second and third harmonics of the 
primary feature {\it and} that we have correctly 
identified their alias orders.  This conclusion 
is further corroborated by the fact that 
$1/P_2(\rm ms)$ is about $N/P_1(\rm ms)$, where 
$N$ is the peak harmonic number (35) associated 
with the primary modulation feature. 

In summary, the above analysis has assisted us 
in {\it unfolding} 0943+10's otherwise multiply 
folded fluctuation spectrum and ruling out various 
combinations of $P_3$ (along with their implied 
drift directions). The ambiguity, however, between 
the remaining two possible combinations---namely, 
$P_3 < 2 P_{1}$ (thus implying negative drift) and 
$P_3 > 2 P_{1}$ (positive drift)---cannot be 
resolved through the above analysis, owing to
the limited sampling of the modulation (which our 
sightline allows) during each rotation of the star.

\section*{IV. Spacing of the Rotating Subbeams}

Given the regularity of subpulse drift in the 
current 0943+10 sequence, and the implied 
stability of the drift bands, an important 
geometrical issue can offer useful insight.  
If the stable subpulse drift corresponds to a 
circular pattern of rotating subbeams---as is 
envisioned by the received ``cartoon'' of 
pulsar emission as well as by the Ruderman \& 
Sutherland 1975) theory---then we should be 
able to estimate the angular (magnetic azimuthal) 
separation between the adjacent subbeams using 
either a) the observed longitude interval between 
subpulses ($P_2$) or b) the associated rotation 
in polarisation angle (PA)\footnote{0943+10's 
subpulses exhibit no distinct PA signature as 
reported for B0809+74 (Taylor \etal\ 1971) or 
B2303+30 (Gil 1992)---rather, the PA varies 
simply with the longitude and polarisation 
mode.}. Of the two, we can use the PA-rotation 
approach immediately---at least approximately. 

We are then interested in the amount of PA 
traverse within an interval of longitude 
corresponding to the separation between 
subpulses---that is, $P_2$.  If the subpulse 
drift does correspond, in actuality, to a 
pattern of subpulse beams which pivot about 
the magnetic axis, then we might expect that 
the orientation of their linear polarization 
(which is ostensibly fixed to the projected 
magnetic field direction) would correspond 
closely to their magnetic azimuth.  

\begin{figure*}
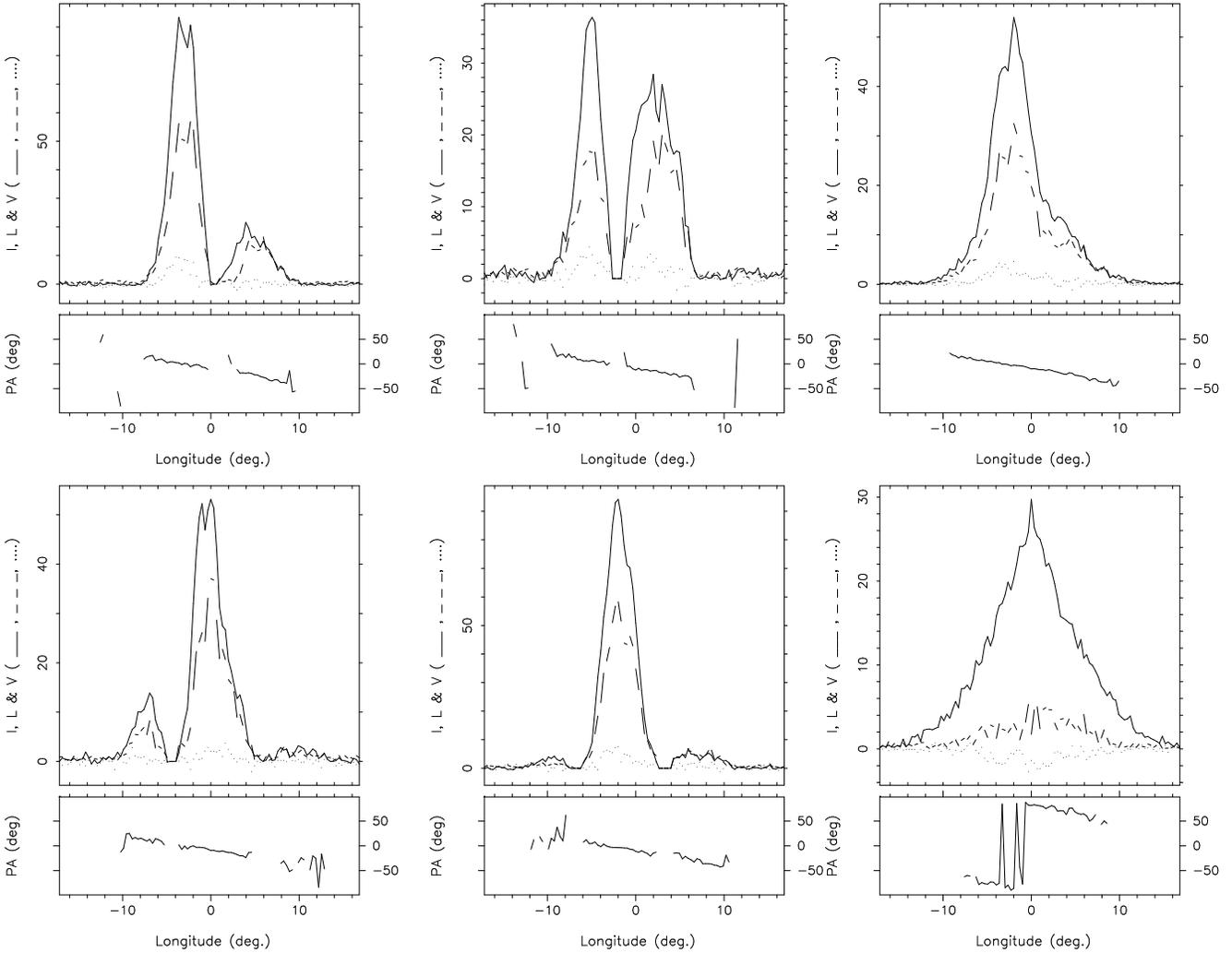

\begin{center}
\begin{tabular}{@{}lr@{}lr@{}}
{\mbox{\epsfig{file=PQmnshort_fig6a.ps,height=5.7cm,angle=-90}}}&
{\mbox{\epsfig{file=PQmnshort_fig6b.ps,height=5.7cm,angle=-90}}}&
{\mbox{\epsfig{file=PQmnshort_fig6g.ps,height=5.7cm,angle=-90}}}\\
{\mbox{\epsfig{file=PQmnshort_fig6c.ps,height=5.7cm,angle=-90}}}&
{\mbox{\epsfig{file=PQmnshort_fig6d.ps,height=5.7cm,angle=-90}}}&
{\mbox{\epsfig{file=PQmnshort_fig6e.ps,height=5.7cm,angle=-90}}}\\
\end{tabular}
\end{center}
%\parhang {\bf Figure 6.}---
\caption{Composite of 4 partial polarisation 
profiles (a,b,d,e) folded at different phases 
of the nominal 1.87- $P_1$ modulation cycle 
depicted in Fig.~\ref{fig:fig6} as well as 
partial profiles of both the ``drifting'' 
power (c) and the ``base'' (f).  All were 
computed from the entire 816-pulse ``B''-mode 
sequence of Fig.~\ref{fig:fig1}.  Two and at 
times three subpulses are seen in the former 
which ``drift'' progressively from 
later to earlier phases about 2--3$\deg$ 
from plot to plot.  Note that the PA
difference between subpulse peaks is 
negative and just less than $30\deg$---a 
circumstance which is just compatible 
with 20 subpulse beams and an inside 
(poleward) sightline traverse (see 
text).  This is very clearly seen in 
the PA behaviour of the ``drifting'' (c)
and ``base'' (f) profiles, where we find 
PA rates of -2.7 and perhaps about -4 
$\deg/\deg$, respectively.  
\label{fig:fig6}}
\end{figure*}

\begin{figure}
\epsfig{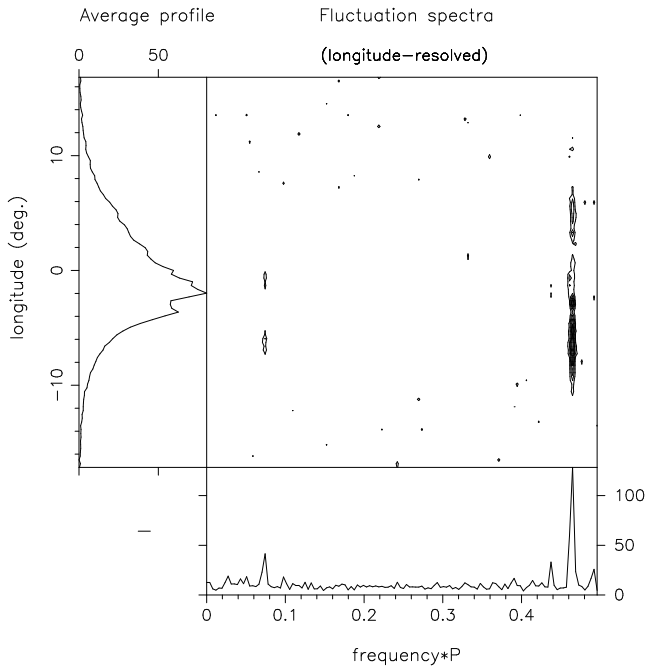}
%\parhang {\bf Figure 7.}---
\caption{Fluctuation spectra as in
Fig.~\ref{fig:fig1} for pulses 129--384.
The ``sidebands'' of the principal feature
are particularly strong in this sequence
(see text).  The contour interval is 102
mJy$^2$ up to a maximum of 714 mJy$^2$.
\label{fig:fig7}}
\end{figure}

Figure~\ref{fig:fig6}(a,b,d,e) gives a 
composite of four average Stokes profiles 
corresponding to equal intervals of phase 
within the 1.87-$P_1$ primary modulation cycle.  
As expected, we see the average characteristics 
of two (or sometimes three) subpulses at 
progressively varying positions in longitude.  
These figures delineate the average polarisation 
of the drift sequence, and we will return to 
a fuller discussion of them below.  However, 
note that the rotation of the polarisation 
angle (PA) over an interval corresponding to 
the spacing between the centres of these 
averaged subpulses ($P_2$)---what we call 
$\chi_{P_2}$---is about $-28\deg$ when the 
subpulse pair straddles the longitude origin 
(or, in the ``drifting'' profile, Fig.~\ref{fig:fig6}e).  
This immediately provides a direct means of 
estimating the number of subbeams which are 
possibly arranged along a circular ring centred 
on the magnetic axis. Neglecting spherical 
effects and assuming that the subpulse beams 
fall closer to the magnetic than to the 
rotational axis, we find two possibilities:  
if $|\chi_{P_2}| > P_2$($\deg$), the magnetic 
azimuth interval between the subpulse beams 
$\eta$ is $|\chi_{P_2}|-P_2$ and an inside 
(negative, poleward) sightline traverse 
is indicated, whereas if $|\chi_{P_2}| < P_2$, 
then $\eta \sim |\chi_{P_2}|+P_2$ and the 
sightline makes an outside (positive, 
equatorward) traverse\footnote {These 
approximations will be justified in \S VI}.  
Here, $\eta \sim 28\deg - 10\deg \sim 18\deg$, 
and thus $360\deg/18\deg$ indicates about 20 
subbeams.  A more precise determination of 
this number should be possible once the 
overall geometry has been determined, or, 
on the basis of some other consideration (as 
we will see in the following section).  It 
follows that the system of such subbeams 
would make a full rotation around the 
magnetic axis in an interval of just 20 
times $P_3$---that is, around  40 rotational 
periods of the star.

\section*{V. The Modulation on the Modulation}

Given the unprecedented stability of the phase 
modulation in the 0943+10 PS discussed above, 
we can legitimately attempt a level of analysis 
which would heretofore have been inconceivable. 
It is clear that any underlying pattern of 
rotating subbeams is stable over many times 
the above 40-$P_1$ scale.  This immediately 
implies that their number must be an integer. 
Also, we would expect to see a periodic feature 
corresponding to, in general, the rotational 
cycle of the subbeam pattern unless all the 
subbeams are identical (or utterly random) in 
amplitude.  Such a tertiary periodicity, if 
any, would manifest itself as sidebands to the 
spectral features associated with the primary 
phase modulation.  Whether such sidebands are 
symmetric or not would depend on whether the 
subbeams differ from each other only in intensity, 
or in interbeam spacing or in both, and their 
narrowness would depend on how stable the 
subbeam pattern is in time.  

Two other minor features can be discerned in
Figs.~\ref{fig:fig1} and \ref{fig:fig4} which
fall symmetrically about the primary one, and
their change of position in Fig.~\ref{fig:fig4}
shows that all three features are aliased responses
in Fig.~\ref{fig:fig1}.  These features are seen
even more clearly in the spectra corresponding
to a small part of the sequence (pulses 129--384)
as shown in Figure~\ref{fig:fig7}.  Careful
measurements (using the fluctuation spectra over
only a few degrees of central longitudes where the
modulation sidebands are significantly stronger)
fall at frequencies about $0.0268\pm0.00037$
c/$P_1$ higher and lower than does the primary
one at 0.535 c/$P_1$.

The separation of the minor feature pair from 
the primary is indicative of a major frequency 
component in the fluctuation power of the tertiary 
amplitude modulation.  Lower frequency amplitude 
modulation would produce symmetrical features 
closer to the fundamental, and nothing is seen 
in the fluctuation spectra here.  Consequently, 
that there is just one obvious such pair of 
``simple''---that is unimodal---features associated 
with the tertiary amplitude modulation suggests 
that it is itself here tone-like.  Were this 
tertiary modulation more complex, we would expect 
to see additional pairs of symmetrical features 
at larger (but harmonic) separations from the 
primary one and, only a slight deviation from 
the tone-like modulation is discernible in 
Fig.~\ref{fig:fig7}.  

As already noted, the ``drifting'' character 
of 0943+10's pulse sequences as well as the 
asymmetrical nature of the primary feature in 
its fluctuation spectrum are strongly indicative 
of a phase modulation. The {\it narrow} and 
{\it symmetrical} pair of features about the 
primary one in Fig.~\ref{fig:fig7} (and also in 
Figs.~\ref{fig:fig1} \& \ref{fig:fig4}) strongly  
suggest the presence of a regular, highly periodic, 
amplitude modulation {\it of} the exquisitely 
regular phase modulation associated with the 
star's ``drifting''-subpulse sequences.  This 
suggests that the putative subbeams differ from 
each other primarily in their intensities, and 
that the intensity variations are a rather smooth 
and stable function of magnetic azimuth.   

The period of this tertiary modulation is then 
about 1/0.027 $P_1$/c or just over 37 rotation 
periods.  We note immediately that this, in turn, 
is very close to 20 times the period of the 
fundamental phase modulation, which at 1/0.5355 
$P_1$/c or some 1.867 rotation periods, has a 
20-cycle period of some $37.35\pm0.03$ rotation 
periods.  This agreement is no accident or 
coincidence; any tertiary modulation not fully 
commensurate with the primary phase modulation 
would either broaden the primary feature or 
distort the ``sidelobes'' around it.  Given 
the accuracy to which the frequencies of the 
secondary and the tertiary modulations can be 
estimated, we can now conclusively rule out 
the possibility that $P_3$ is greater than 2 
rotation periods, because 1/0.4645  $P_1$/c is 
simply not harmonically related to 37.35  $P_1$.  

The period of this tertiary modulation is then
about 1/0.027 $P_1$/c or just over 37 rotation
periods ({\it i.e.} $37.3\pm0.5$). Given that 
the length of the sequence analysed corresponds 
to many times this tertiary cycle and that the 
``sidelobe'' features have the sharpness suggestive 
of a remarkably stable modulation, an harmonic 
relationship between the two periods must exist.
We note immediately that tertiary modulation cycle
is indeed very close to 20 times the period of the
fundamental phase modulation, which at 1/0.5355
$P_1$/c or some 1.867 rotation periods, has a
20-cycle period of some $37.35\pm0.03$ rotation
periods.  This agreement is no accident or
coincidence; any tertiary modulation not fully
commensurate with the primary phase modulation
would either broaden the primary feature or
distort the ``sidelobes'' around it.  Given the 
accuracy to which the frequencies of the secondary 
and tertiary modulations can be estimated, we can 
now conclusively rule out the possibility that 
$P_3$ is greater than 2 rotation periods, because 
1/0.4645 $P_1$/c is simply not harmonically related 
to 37.35 $P_1$.  The only commensurate value of 
$P_3$ (the one smaller than 2 $P_1$, as discussed
earlier) implies with certainty that the drift 
direction is {\it negative}.  The observed harmonic 
relationship then also allows us to now adopt a 
refined estimate of the tertiary modulation period
as simply 20 times the better-determined value of 
$P_3$.  It is important to mention here that, while 
the integral relationship between the modulation 
frequencies has resolved the longstanding 
drift-direction ambiguity, the exact number of 
subbeams (20) is in no way crucial to our further 
analysis below.

Therefore, we believe that this analysis has resolved 
the question of how and which way pulsar 0943+10's 
subpulses drift.  Sieber \& Oster's (1975) fig. 1 
delineates some of the 
aliasing possibilities. The ``A'' band represents an 
apparent drift with a $P_3$ of some 15, and it seems 
to entail gradual subpulse motion from later to 
earlier phases---that is, a {\it negative} drift.  
Band ``B'' ostensibly represents a negative drift 
associated with a value of $P_3$ just less than 2, 
and Band ``C'' one with a $P_3$ just greater than 2. 

Stated differently, Sieber \& Oster's diagram clearly 
indicates the nearly alternate-pulse character of the 
basic modulation.  So, it is not straightforward to 
distinguish whether the basic drift is to the right 
or to the left.  Furthermore, 
as these authors emphasize, the LRF spectra in 
Fig.~\ref{fig:fig1}---and all others like it---have 
the same difficulty.  Any of the features in such 
a diagram can be aliases of fluctuations 
at frequencies greater than the 0.5 c/$P_1$ Nyquist 
frequency associated with sampling at a $1/P_1$ rate.  
Only by appealing to the finite width of the pulse 
can this ambiguity be alleviated as we have 
demonstrated above. 

The ``B'' track in Sieber \& Oster's plot is then
the fundamental oscillation 
associated with 0943+10's drift, and the ``A'' track 
its first harmonic.  This establishes the drift 
direction as negative.  The fundamental 0.535 
c/$P_1$ periodicity implies an associated $P_3$ 
value of some 1.867 periods---that is, in two 
rotations we see the {\it next} subpulse at a 
slightly earlier phase.  Also, the effective ``$P_3$'' 
of the second harmonic is $1/1.07$ or about 
0.93 $P_1$/c---whose difference from 1.0 $P_1$/c 
describes the ``A`` track, while the conspicuously 
missing alternate subpulses are due to the 
fundamental modulation. 

\begin{figure}
\epsfig{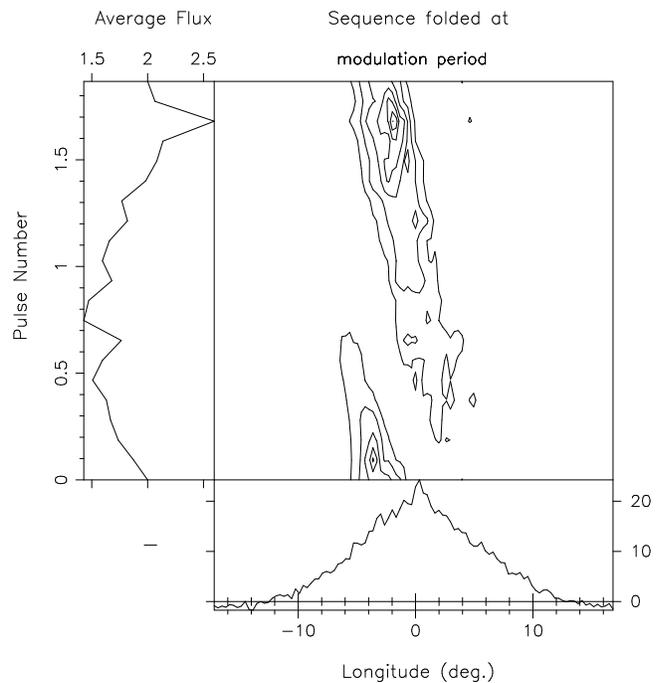}
%\parhang {\bf Figure 8.}---
\caption{Block-averaged pulse sequence 
which corresponds to the principle ``drift'' 
pattern in pulsar 0943+10's ``B'' mode.  The 
816 individual pulses are folded at the 
1.867-rotation period interval, which in 
turn corresponds to the primary modulation 
feature (alias resolved) at some 0.5355 
c/$P_1$ in Fig. 2.  The strength of the 
feature as a function of longitude and phase 
within the 1.87-period cycle is given as a 
contour plot in the body of the figure, and 
its integrated energy as a function of phase 
in the right-hand panel, whereas the 
aperiodically-fluctuating ``base'' (which 
has been removed from the other displays 
prior to plotting) is given in the bottom 
panel.  Contours fall at about 26.7 mJy up 
to a maximum of 187 mJy.  
\label{fig:fig8}}
\end{figure}

The basic character of this drift behaviour 
can now be unambiguously represented by folding 
the entire 816-pulse ``B''-mode sequence at the 
1.867-$P_1$ interval corresponding to the 
primary 0.535 c/$P_1$ feature.  We give such a 
display in Figure~\ref{fig:fig8}, where the 
subpulse intensity as a function of longitude 
and phase within the modulation cycle is shown 
as a contour plot in the central panel.  The 
varying energy in the modulation is given in 
the right-hand panel and the non-fluctuating 
``base'' (minimum-intensity) profile is shown 
in the bottom panel.  While this display 
resolves longstanding questions about pulsar 
0943+10's ``drift'', we now also see that its 
overall character is remarkably complex.  

Finally, the conclusion that 0943+10's ``drift'' 
is {\it negative} is very important to arguments 
we will make in the following sections.  The 
actual physical direction of the subpulse 
``drift'' {\it observed} is negative---that is, 
it is apparently in the {\it same} direction as 
the pulsar's rotation.  

Having established that the tertiary 
modulation cycle is 20 times the primary $P_3$ 
period of 1.867 $P_1$/c, the interpretation 
is clear.  We therefore conclude that 
the observed ``drifting'' behaviour is a 
manifestation of the rotation of a relatively 
stable system of 20 subbeams  of systematically 
differing intensity, spaced uniformly in 
magnetic azimuth.  

\begin{figure}
\epsfig{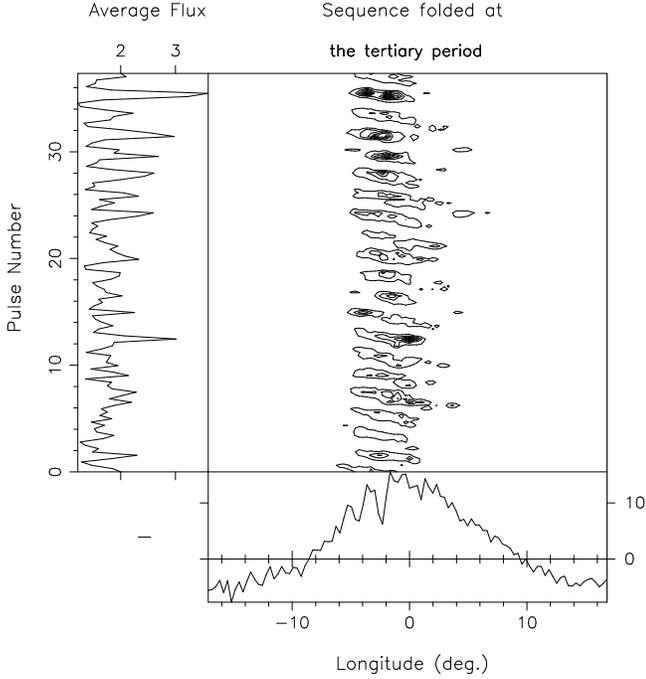}
%\parhang {\bf Figure 9.}---
\caption{Block-averaged pulse sequence
corresponding to the 430-MHz ``B''-mode
observation in Figs.~\ref{fig:fig1}--\ref{fig:fig4}.
The 816 individual pulses are folded at
the tertiary modulation period, which
was calculated to be 37.346 rotation
periods.  Average pulse intensity as
a function of longitude is represented
in the body of the figure as a contour
plot (contours at 51 mJy up to 457 mJy).
The longitude-integrated intensity over
the period of the tertiary-modulation
cycle is plotted in the left-hand panel,
and the amplitude of the
aperiodically-fluctuating ``base''---which
is subtracted from the other displays---is
given in the bottom panel.
\label{fig:fig9}}
\end{figure}

One means of verifying this circumstance is to 
fold the pulse sequence at the period of the 
tertiary modulation, and it would be prudent to 
use the value of this period based on the phase 
modulation, which we calculated above.  The 
result for the 816-pulse sequence under 
consideration is shown in Figure~\ref{fig:fig9}, 
and it bears detailed discussion.  The average 
amplitude as a function of longitude over the 
tertiary-modulation cycle is depicted in the 
body of the figure; the integral over longitude 
is given at the left, and the amplitude of the 
aperiodically-fluctuating ``base'' (which has 
been subtracted before plotting the other 
panels) is shown in the bottom panel.\footnote
{We note here that this ``base'' profile (see 
also Fig.~\ref{fig:fig6}(f) has a width of 
some 11--$12\deg$.  It is not at all clear what 
interpretation should be made of this ``steady'' 
(as against the phase-modulated) contribution 
to the pulse sequence, but, were we to interpret 
this ``base'' feature as a core component, it 
is interesting both that its form is symmetrical 
about the longitude origin (whereas, the 
``drifting'' power is not at all so) and its 
width is suggestive of a magnetic colatitude 
of some $11\deg$--$12\deg$ [Rankin (1990): 
eq.(5)].} 

First, note the remarkable circumstance that 
folding at the tertiary-modulation period reveals 
an orderly subsequence of 20 replicating emission 
elements, each of which have their own distinct 
relative amplitude and behaviour as a function 
of longitude.  As is expected, one of the 20 
elements is significantly stronger than the 
others by upwards of a factor of two.  Although 
each extends over $10\deg$ or more in longitude, 
some peak much earlier than others and exhibit 
a surprising range of individual characteristics; 
the brightest one, for instance, has a double 
amplitude structure.  We emphasize that these 
elements {\it must} be relatively stable over 
this 15-minute ``B''-mode sequence; it is their 
consistent brightness, and systematic variation 
in brightness, which apparently produces the 
amplitude modulation at a frequency $1/20$-th 
that of the primary phase modulation.  We do 
not usually think of the processes associated 
with subpulse ``drift'' as having stability on 
such long time scales---and it is very possible 
that they usually do not.

Nonetheless, given the regularity of subpulse
drift in the current 0943+10 sequence and the
evidence that it entails a further modulation
indicative of a repeating pattern of just 20
elements, we are forced to reconsider received
ideas about the ``drifting subpulse phenomenon''
and its possible origin in a pattern of slowly
rotating subpulse beams (Ruderman \&  
Sutherland 1975).  Our results provide 
unexpectedly strong support for the simple 
``cartoon'' understanding of ``drifting'' 
subpulses as a regular pattern of emission 
centres rotating around the pulsar's magnetic 
axis. 

\section*{VI. Pulsar 0943+10's Emission Geometry}

We expect a pulsar with {\it drifting} subpulses 
to have an emission geometry characterized by a 
highly tangential traverse through (or, perhaps 
better, along) one of the hollow conical beams 
associated with pulsar radiation.  And, uniquely 
for 0943+10, our sightline apparently grazes its 
beam so narrowly that we miss ever more of its 
emission at higher frequencies.  The pulsar is 
thus an excellent example of the conal single 
({\bf S}$_{\rm d}$) class (Rankin 1993{\it b}), 
and its value of $B_{12}/P^2$, which is about 
unity, supports this classification.  Questions, 
nonetheless, remain about its geometry:  Do we 
observe an inner or an outer cone ({\it e.g.},
Rankin 1993{\it a})?  Does our sightline pass 
poleward or equatorward of the magnetic axis 
({\it e.g.}, Narayan \& Vivekanand 1982)?  Can 
we understand the pulsar's steep spectrum in 
terms of the conal narrowing with frequency?  
And finally, is the overall conal geometry 
consistent with the 20-beam ``carousel'' 
pattern indicated by the tertiary modulation?  

Information on the pulsar's geometry comes from 
considering its PA traverse in conjunction with 
its profile width.  Then, we can model these values 
using the relations in Rankin (1993{\it a}), eqs. 
(1-6).  Reference to Suleymanova \etal's fig. 4---which 
treats the same 1992 430-MHz sequence---shows the 
PA behaviour for the 816 ``B''-mode and 170 
``Q''-mode pulses, and it is clear that the two 
modes exhibit somewhat different PA sweep rates. 
Their Table II gives values of $-2.4\deg/\deg$ 
and $-3.6\deg/\deg$ for the ``B'-' and ``Q''-mode 
PA rates at 430 MHz and $-3.4\deg/\deg$ and 
$-3.6\deg/\deg$ at 102.5 MHz, respectively.  
Also important is their fig. 8, which shows the 
evolution of the ``B''- and ``Q''-mode profiles 
between 25 and 430 MHz.  A further set of time-aligned 
profiles from Arecibo observations is given by 
Hankins, Rankin \& Eilek (2000).  

Our initial problem for 0943+10 is that we do 
not know where we cross its 430-MHz beam---that 
is, where $\beta$ (the impact angle of the 
sightline with respect to the magnetic axis) 
falls with respect to $\rho$ (the outside, 
half-power radius of the conal emission beam).   
At higher frequencies our sightline cuts well 
outside of the radial maximum and ultimately 
outside the half-power point $\rho$, whereas 
at lower frequencies the double profile forms 
indicate cuts inside the radial maximum point.  
Therefore, the observed dimensions of the 
430-MHz profile will probably not correspond 
to those given for the overall pulsar population 
by Rankin (1993a)---even if we could extrapolate 
properly to 1 GHz---because those are based on 
more central sight-line traverses that {\it do} 
cross the radial maximum point on the hollow 
conical emission beam.  

Let us then model the 0943+10 geometry for each 
of four possible configurations (inside traverse, 
inner cone; inside traverse, outer cone; outside 
traverse, inner cone; and outside traverse, 
outer cone) as follows:  a) Calculate the conal 
beam radii $\rho_{ \rm outer(inner)}$throughout 
the entire frequency range using Mitra \& 
Deshpande's (1999) expression [patterned after 
Thorsett's (1991) profile-width relation] with 
trial values of the index $a$
$$\rho = \rho_{\rm outer(inner)}{(1 + 0.066 f_{\rm GHz}^{-a}) P_1^{-{1\over 2}}\over 1.066}, \eqno(1)$$
where the 1-GHz values of the conal beam radii 
$\rho_{\rm outer(inner)}$ are taken according to 
Rankin's eq. (5)\footnote{While most pulsars 
exhibit what Rankin (1993{\it a,b}) defined as 
``inner'' and ``outer'' cones, there is now 
evidence that  a few stars exhibit cones with 
both larger and smaller characteristic radii 
(Gil \etal\ 1993; Kramer \etal\ 1994; Rankin \& 
Rathnasree 1997; Mitra \& Deshpande 1999).}.  
b) Least-squares fit the observed low frequency 
profile widths (111.5 MHz and below) to calculated 
values from the usual spherical geometric relation
$$\Delta\psi = 4 \sin^{-1}\left[{{\sin(\rho/2+\beta/2) \sin(\rho/2-
\beta/2)}\over{\sin\alpha\sin(\alpha+\beta)}}\right]^{1\over 
2}\eqno(2)$$
(Gil 1981; Rankin 1993{\it a}) with trial values 
of the magnetic co-latitude $\alpha$ and $\beta$. 
c) Least-squares fit the frequency-dependent part 
of eq. (1) to obtain the index $a$, using it 
in step a) above in an iterative solution.  With 
this procedure, one is in a position to compute 
the values of $\beta/\rho$ even at frequencies 
where $\beta$ exceeds $\rho$.  

Two further useful quantities can be calculated 
here. One is the magnetic azimuth angle $\eta$ 
that corresponds to a particular value of $P_2$.  
Using very simple spherical geometry, this angle 
is given (for small angles) by 
$$\eta = 2 \sin^{-1}[\sin(P_2/2)\sin(\alpha+\beta)/|\sin{\beta}|], \eqno(3)$$  
where $P_2$ is measured as a longitude interval 
(not a time), and, for 0943+10 at 430 MHz has a 
value of just over $10\deg$.  Reference to 
Fig.~\ref{fig:fig6} gives $P_2$ values in the 
8.5-9.0$\deg$ range, but a more accurate value 
comes from the fluctuation-spectral phase rate 
of the primary (0.535 c/$P_1$) feature in 
Fig.~\ref{fig:fig5}---around $0\deg$ longitude 
the phase rate is some $34\pm1\deg/\deg$, which 
implies a $P_2$ value of $10.6\pm0.6\deg$.  [The 
smaller values implied by Fig.~\ref{fig:fig6} 
probably reflect the curvature of the subbeam 
track relative to the sightline.  Note that (see 
just below) $10.6\deg \cos{\chi_{P_2}} \approx 
9\deg$].  Clearly, if the drifting represents a 
pattern of 20 subbeams around the periphery of 
the polar cap, then this angle $\eta$ must take 
a value of just $18\deg$.  

Another geometrical parameter associated with  
the drifting pattern is the PA rotation between 
subpulses---what we called ``PA $P_2$'' above.  
This PA rotation directly reflects the relative 
curvatures of the sightline and the conal beam.   
Its value is easily calculated from the usual 
single-vector model (Radhakrishnan \& Cooke 1969;
Komesaroff 1970) by evaluating 
$\chi_{P_2} = \chi(P_2/2) - \chi(-P_2/2)$, or 
$$\chi_{P_2}=2\tan^{-1}[{{\sin\alpha \sin(P_2/2)}\over{\sin(\zeta) \cos\alpha-\cos(\zeta) 
\sin\alpha \cos(P_2/2)}}], \eqno(4) $$
where $\zeta = \alpha + \beta$.  Incidentally, 
eqs.(3) and (4) reduce nicely to rationalize the 
rules we used in \S IV.  For small angles eq.(4) 
reduces to $\chi_{P_2} \simeq R_{PA} P_2$ (where 
the PA sweep rate $R_{PA} = \sin\alpha/\sin\beta$), 
and eq.(3) to $\eta \simeq (|R_{PA}| + |\beta|/\beta) P_2$; 
thus $\eta \approx \chi_{P_2} + {\rm sgn}\beta P_2$ 
and will always be positive as long as the subbeams 
do not fall closer to the rotation axis than to the 
magnetic axis.  

Computations for all four emission geometry models 
were carried out to determine the best-fitting 
values of $\alpha$, $\beta$, and $a$, such that 
$\chi^2 = (1/3)\Sigma [(w - w_c)/\Delta w]^2$, where 
$\Delta W$ was taken as 0.4$\deg$.  The first of 
these models---our preferred one, for an inner 
traverse and inner cone---is given in Table II.  
The columns giving $\rho$, $\beta/\rho$, and the 
emission height $h$ [from the centre of the star; 
see Rankin's (1993{\it a}) eq.(6)] depend only on 
the fitted values of $\alpha$, $\beta$, and $a$.  
$w$ and $w_c$ are the respective measured (taken, 
wherever possible in the ``B'' mode for consistency) 
and modeled outside, half-power profile widths 
[the latter from eq.(2)].  $\rho_c$ (not shown) 
was computed for comparison from Rankin's eq.(4), 
and the quantities $R_{PA}$, $\eta$, and $\chi_{P_2}$ 
evaluated from the best-fitting parameters.  

A quick scan of Table II shows immediately that 
the inner cone, inner traverse model yields a 
very good fit and reasonable parameters.  This 
is also true for the outer cone, inner traverse 
model which gives almost identical results, but 
with $\alpha$ 15.39$\deg$ and $\beta$ -5.69$\deg$.  
The two models have virtually identical values of 
$\beta/\rho$, $R_{PA}$, $\eta$, $\chi_{P_2}$, and 
$\chi^2$, so the fitting does not distinguish 
between them; indeed, it appears that models with 
any reasonable 1-GHz $\rho$ value will behave 
similarly.  

By contrast, the outer traverse models are of two 
different types:  Inner and outer cone models with 
values of $\alpha$ and negative $\beta$ similar to 
those above have nearly identical values of $R_{PA}$ 
and $\chi_{P_2}$ (and $a$ about 0.53), but yield 
slightly poorer fits ($\chi^2 \sim 2.2$).  These, 
however, result in very different values of 
$\eta$---about 42$\deg$---as might have been expected 
from eq.(3), which depends strongly on the sign of 
$\beta$, unlike the various other geometrical 
relations [{\it i.e.}, eqs.(2) or (4)].  Finally, 
there are a pair of inner and outer cone ``pole 
in cone'' models with $\alpha$ about 2.9 and 
4.0$\deg$, respectively, and $R_{PA}$ 0.7, $\eta$ 
18$\deg$, $\chi_{P_2}$ about +7.4$\deg$, and 
$\chi^2$ some 1.6.  Clearly, none of these models 
are acceptable because the former fail to match 
the observed values of $\eta$ and the latter 
those of $R_{PA}$ and $\chi_{P_2}$.  

\begin{center}
Table II. Inner Cone, Inner Traverse Emission Model   
\begin{tabular}{|c|c|c|c|c|c|}\hline\hline
f(GHz) & $\rho$($\deg$) & $\beta/\rho$ & h(km) & w($\deg$) & $w_c$($\deg$) \\ \hline
1.000 & 4.13 & -1.04 & 125 & ---  & ---  \\
0.430 & 4.23 & -1.01 & 131 & 7?   & ---  \\
0.306 & 4.29 & -1.00 & 134 & ---  &  0.1  \\
0.111 & 4.49 & -0.96 & 147 & 16.5 & 16.7 \\
0.103 & 4.51 & -0.95 & 149 & 17.5 & 17.5 \\
0.061 & 4.65 & -0.92 & 158 & 23   & 22.7 \\
0.049 & 4.72 & -0.91 & 163 & 25   & 24.8 \\ 
0.040 & 4.79 & -0.89 & 168 & 27   & 26.9 \\
0.034 & 4.85 & -0.88 & 172 & 28   & 28.6 \\
0.025 & 4.98 & -0.86 & 181 & 32   & 31.8 \\ \hline 
\end{tabular}
{\it Observational constants}: $P_2$, 10.5$\deg$ \\
{\it Fitted parameters}: $\alpha$, 11.58$\deg$; $\beta$, -4.29; $a$, 0.396; $\chi^2$, 1.12 \\
{\it Derived parameters}: $R_{PA}$, -2.7$\deg/\deg$; $\eta$, 18.0$\deg$; $\chi_{P_2}$, -27.9$\deg$
\end{center}

Our slightly circuitous means of computing pulsar 
0943+10's geometry---using only the profiles at 111 
MHz and below with resolved double forms indicative 
of $|\beta|/\rho$ well less than unity---permits 
us to assess how the sightline cuts the pulsar's 
conal beam at higher frequencies.  The row in Table 
II just above 111 MHz gives the frequency for which 
$w_c$ is near zero---that is, the frequency at which 
the model $\rho$ would be just equal to $|\beta|$.  
At 430 MHz, the table shows that our sightline passes 
just outside the half-power radius on the conal beam, 
by $0.07\deg$ and $0.09\deg$, or just less than 2\% 
according to the two inner-traverse models, 
respectively.  No wonder that 0943+10 is so 
difficult to detect above 400 MHz!

We can now draw some remarkable conclusions.  We 
noted in \S III that 0943+10's subpulse drift is 
negative---that is, in the same direction as its 
rotation.  Then, we concluded just above that our 
sightline makes an inside (negative) traverse---that 
is, crossing between the magnetic and rotational 
axes, with a negative rotation of the PA.  If we 
imagine that we are looking at the pulsar with the 
``closer'' rotational pole ``up'', the emission 
beams sweep past our sightline from right to left, 
thus implying that from our perspective, the pulsar 
is rotating clockwise.\footnote{Our discussion here 
is not fully consistent with the conventions of the 
rotating-vector model, which are reviewed nicely by 
Everett \& Weisberg (2000).  In our case, the 
``closer'' rotational pole points away from the 
direction of the star's angular momentum vector, so 
that $\alpha_{\rm RVM}$ is 180$\deg$ - $\alpha$, 
and $\beta_{\rm RVM}$ is $-\beta$}.  

Further, the subpulse beams are also rotating from 
right to left, but around the magnetic axis, which 
is ``below'' us; therefore, they rotate around the 
polar cap counterclockwise (in the frame of the 
star).  This drift in 0943+10 is opposite to the 
Ruderman \& Sutherland (1975) sense (see their fig. 
4); they discuss Ferraro's Theorem in an Appendix 
but do not clearly state its implication that polar 
cap plasma must rotate in the same direction as 
the star as viewed by an inertial observer.  
Ruderman (1976) gives a correct and clear 
discussion of this point.

\section*{VII. Imaging the Emission Zone}

The remarkable order implied by Fig.~\ref{fig:fig9} 
of just 20 subpulse emission beams rotating about 
the star's magnetic axis demands a further level of 
analysis.  We have therefore developed a coordinate 
transformation between the usual system of pulsar 
(rotational) colatitude $\zeta$ ($ = \alpha+\beta$) 
and longitude $\varphi$ and a system rotating about 
the magnetic axis described by colatitude $R$ and 
azimuth $\Theta$.  The emission beams rotate around 
the pole with a period $\hat{P_3} = N P_3$, where 
(just as in Ruderman \& Sutherland 1975) $N$ is the
 number of subbeams and $P_3$ the usual subpulse-drift 
interval.  

If, further, we number the pulses by $k$ from a 
reference pulse $k_0$ and measure the longitude 
$\varphi$ relative to an origin defined by the 
longitude of the magnetic axis $\varphi_0$, then 
$\Theta$ is the sum of a rotation $\theta_{\rm rot}$ 
and a transformation $\theta_{\rm trans}$ as follows 
$$ \Theta = -\theta_{\rm rot} + \pm\theta_{\rm trans}\ , \eqno(5) $$
where the sign of $\theta_{\rm rot}$ will always be 
negative according to Ferraro's theorem, and the sign of 
$\theta_{\rm trans}$ is positive for cw rotation 
(of the rotational pole which is closest to the 
sightline) and negative for ccw rotation.  Then 
$$\theta_{\rm rot}={2\pi}[k-k_0+(\varphi-\varphi_0)/{2\pi}]/\hat{P_3}, \eqno(6) $$ 
$$\theta_{\rm trans}=\sin^{-1}[{\sin\zeta\sin(\varphi-\varphi_0)}/\sin R ]\ , \eqno(7) $$
where 
$$R=2\sin^{-1}[\sin^2(\{\varphi-\varphi_0\}/2)\sin\alpha\sin\zeta+\sin^2 (\beta/2)]^{1\over 2}\ . \eqno(8) $$
%Finally, we are open to the possibility that the 
%rotation of the subbeam pattern is not rigid, but has 
%a dependence on the angular distance from the magnetic 
%axis.  We model the possible $\hat{P_3}$ dependence 
%on $R$ as follows 
%$$\hat{P_3} (R)=\hat{P_3} (\beta) [{R/\beta}]^{\delta_R}\ , \eqno(9) $$
%where $\delta_R$ is, in general, a real number.  
%However, in the present case the apparent constancy 
%of $P_3$ across the pulse window (as seen in 
%Fig.~\ref{fig:fig1}) suggests that $\delta_R = 0$, 
%a provisional value we will assume.  

Correct identification of the longitude of the 
magnetic axis $\varphi_0$ is very important to 
carrying out the above transformation.  Usually, 
the magnetic axis is expected to fall close to a 
profile's center, but for 0943+10 at 430 MHz, we 
have seen in the companion paper that the ``B''-mode 
profile is truncated on its trailing edge, compared 
with its ``Q''-mode counterpart.  It is, however, 
only the ``B''-mode profile that is truncated at 
430 MHz, its ``Q''-mode counterpart is much broader; 
if we take the center of the latter as close to the 
magnetic axis, note that this point falls close to 
the trailing half-power point of the 430-MHz, 
`B''-mode profile (see Suleymanova \etal: figs. 4 
\& 8).  [The inflection point of the PA traverse 
should lie close to the magnetic axis (apart from 
small relativistic effects, Blaskiewicz \etal\ 
1991), but that point is difficult to determine 
accurately for this pulsar.]  Finally, we can 
appeal to the time-aligned average profiles in 
Hankins \etal\ (2000).  Note that some of the 430-MHz 
profiles are narrow and early (and thus presumably 
``B''-mode dominated); whereas another one is more 
extended and appears to be primarily a ``Q''-mode 
average.  Overall, we see here that the point on 
the 430-MHz, ``B''-mode profiles which aligns best 
with the centers of their low frequency counterparts 
is the half-power point on their trailing edge.  We 
will then take this point, provisionally, as our 
best estimate of the longitude of the magnetic axis.  

\begin{figure}
\epsfig{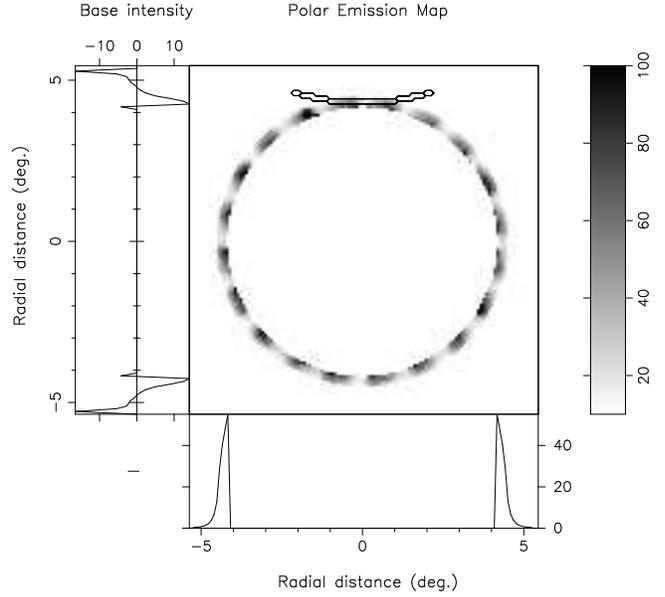}
%\parhang {\bf Figure 10.}---
\caption{An image of the accessible 
emission zone reconstructed using the 
``cartographic'' transform as defined by 
eqs.(5--9). This image is based on the 
observed ``B''-mode sequence of 816 pulses. 
The image or polar map is projected onto 
the polar cap shown in the main panel, 
with the ``closer'' rotational axis at 
the top of the diagram.  The bottom and 
the side panels show the average- and
the ``base''-intensity profiles, 
respectively, as functions of the angular 
distance from the magnetic axis ({\it 
i.e.}, magnetic colatitide).  Contours 
indicate the sightline path at the top 
of the diagram.  Here, the star rotates 
clockwise, causing the sightline to cut 
the counter-clockwise-rotating subbeam 
pattern from left to right.  
\label{fig:fig10}}
\end{figure}

With all the necessary ingredients at hand, we 
proceed to perform the ``cartographic'' transform 
defined by the above equations (and the inside 
traverse, inner cone model in Table II) and to {\it 
reconstruct} an image of the accessible emission 
zone, based on the observed ``B''-mode sequence of 816 
pulses. The result is given in Figure~\ref{fig:fig10}, 
with the emission-pattern image or ``map'' projected 
on to the polar cap shown in the main panel.  The 
bottom and the side panels show the average- and the 
``base''-intensity profiles, respectively, as functions 
of the angular distance from the magnetic axis ({\it 
i.e.}, magnetic colatitide).  The {\it polar map} 
shows the distribution of the 20 nearly uniformly 
spaced subbeams, some noticeably more intense than 
others. The nearly tangential character of our 
sightline traverse is evident from the apparent 
shapes of the individual features, which are almost 
certainly depicting only a small outer portion of the 
``true'' subbeams structure.  The sharply rising 
radial intensity distribution (bottom panel) and the 
absence of any hint of a peak (or an inflection) 
over the accessible range of radii is also 
indicative of this circumstance.  

Note that we have adopted, without loss of generality, 
a simple but useful convention in this transform, 
wherein the rotational longitude associated with the 
magnetic axis defines a vertical line through the 
image centre, with its top pointing to the ``nearer'' 
rotational pole.  So, a negative value of $\beta$, 
as in the present case, would correspond to the 
observer's sightline sampling along a curved track 
through the upper half of the image.  Then, the star's 
rotation is cw when the impact angle $\beta$ has the 
same sign as the PA traverse (with respect to the 
{\it observed} longitude), and ccw when they do not.  
Thus, for the present case, a) the star's rotation 
is in the clockwise direction, and b) the subbeam 
pattern rotates a the ccw direction, given that the 
observed drift is towards the leading edge of the 
pulse window.  

At this point, it is appropriate to ask the 
obvious question:  How unique or even accurate 
are the results given by this ``cartographic'' 
transformation?  The uniqueness---or rather the 
correctness---of the polar map depends only, 
and directly, on the correctness of the 
emission geometry and the other parameters 
which are required to carry out the transform.  
It is very sensitive to the circulation time 
$\hat{P_3}$, the longitude of the magnetic 
axis $\varphi_0$), and the sense of the 
sightline traverse ({\it i.e.}, the sign of 
$\beta$)---and if these quantities are incorrect, 
the image will be smeared or distorted.  The 
actual values of $\alpha$ and $\beta$, however, 
largely just scale the image, so that their 
correct specification is much less crucial.  

In practice, some of these inputs that define 
the transform may not be known with the desired 
accuracy, making this transform less conclusive.  
However, this difficulty can be overcome by 
invoking an ``inverse'' transform, wherein the 
map reconstructed from the original pulse train 
is, in turn, used to produce a new sequence of 
single pulses.  If the parameters used in the 
mapping transformation are indeed {\it correct}, 
then the artificial sequence should match the 
original one in its detailed fluctuation 
properties.  Comparison of the two sequences is 
best effected on a longitude-to-longitude basis, 
such that the new sequence at a given longitude 
is compared with the original one pulse-for-pulse 
and sample-for-sample.  A simple cross-correlation 
coefficient, appropriately normalized, provides 
an adequate quantitative measure for such 
comparison---though, of course, those for each 
range of longitude must be combined through a 
suitably weighted average---indeed, quite a 
robust one, given all the possible ways by which 
the two sequences can differ.  

\begin{figure*}
\begin{center}
\begin{tabular}{@{}lr@{}}
{\mbox{\epsfig{file=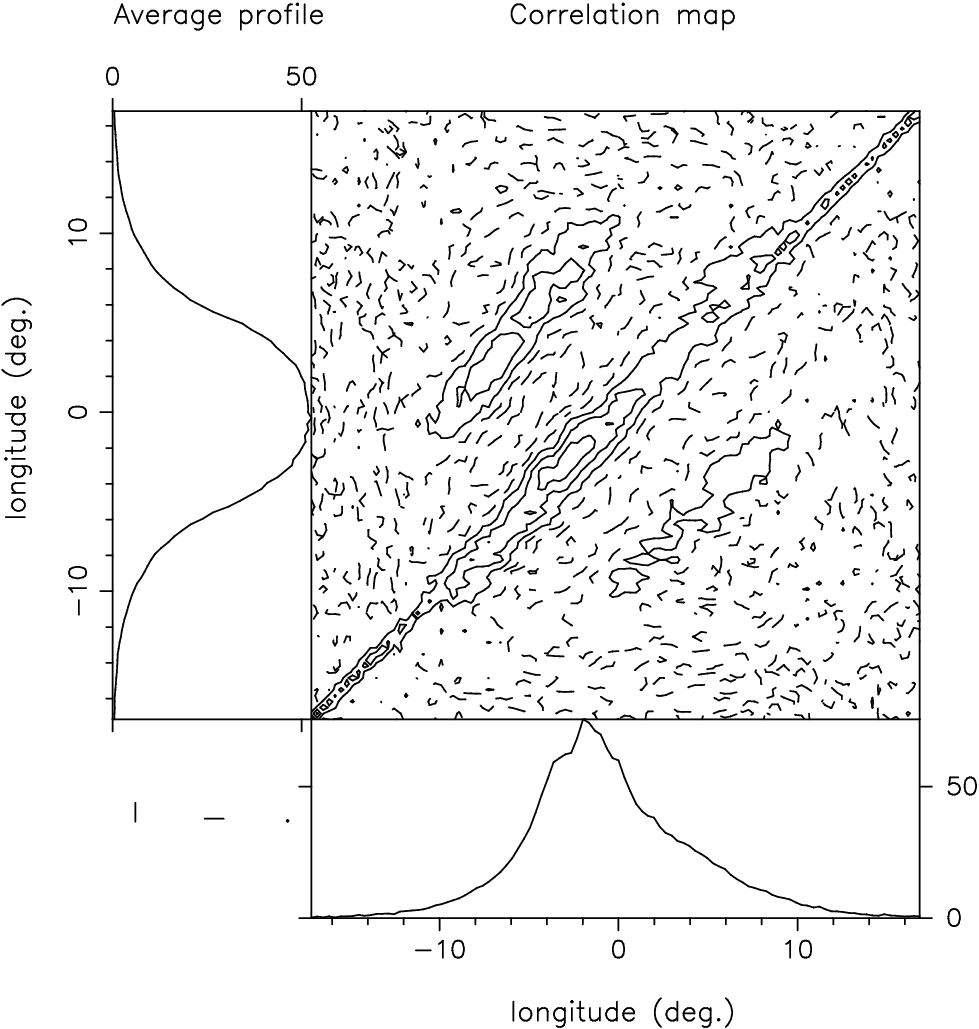,height=7.5cm,angle=-90}}}&
{\mbox{\epsfig{file=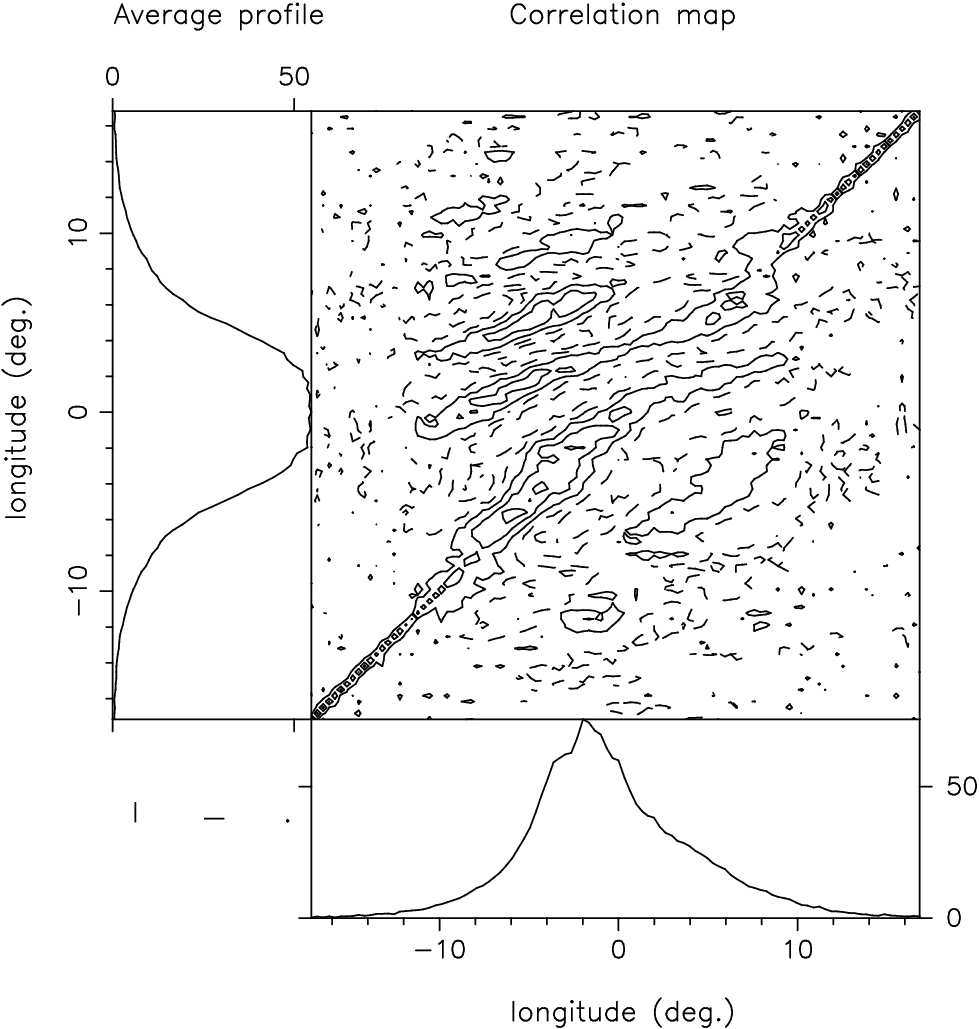,height=7.5cm,angle=-90}}}\\
\end{tabular}
\end{center}
%\parhang {\bf Figure 11.}---
\caption{Cross-correlations between 
the observed pulse sequence and an 
artificial one computed from the 
polar map using the inverse cartographic 
transform.  The central panels display 
the correlation between different longitudes 
across the pulse window.  Two cases are 
shown which differ only in the sign of 
the impact angle $\beta$, illustrating 
the usefulness of the inverse transform 
``closure path'' to assess and refine 
the geometrical parameters on which the 
subbeam imaging depends.  The left-hand 
and bottom panels give the average
profiles corresponding to the artificial 
and the observed sequences, respectively.  
The contour intervals in both plots is 
about 0.12 mJy$^2$.   
\label{fig:fig11}}
\end{figure*}

The inverse transform thus provides a powerful 
``closure'' path to verify and refine the ``input'' 
geometry.  The inverse transform, of course, is defined 
by the same set of relations as for the (forward) 
``cartographic'' transform.  We have used this closure 
path to confirm our present assumptions and also to 
conclusively rule out other geometries that were 
originally considered as plausible.  We illustrate 
this with a simple example of two model geometries, 
differing only in their sign of $\beta$.  Shown in 
Figure~\ref{fig:fig11}{\it a,b} are cross-correlation 
maps of the fluctuations at different longitudes 
across the pulse window.  Plotted in the left and 
bottom panels are the average profiles for the new 
and the original sequences, respectively. It is easy 
to show that the diagonal elements of such correlation 
maps alone would suffice to assess the match.  The 
other details in the displays do carry important 
information and can provide useful clues for 
improving the match.  

However, the most significant potential uses of the 
inverse transform may come through studying what 
features of the observed sequence are lost in the 
process of recreating the artificial sequence from 
the polar-cap map.  Note, for instance, that any
pattern of subbeams will produce a sequence that 
will average to a symmetrical profile---{\it e.g.}, 
see the left-hand panels of Fig.~\ref{fig:fig11}.  
These differ noticeably from the asymmetric 430-MHz 
average profiles observed.

\section*{VIII. The ``Q''-Mode Pulse Sequence }

\begin{figure}
\epsfig{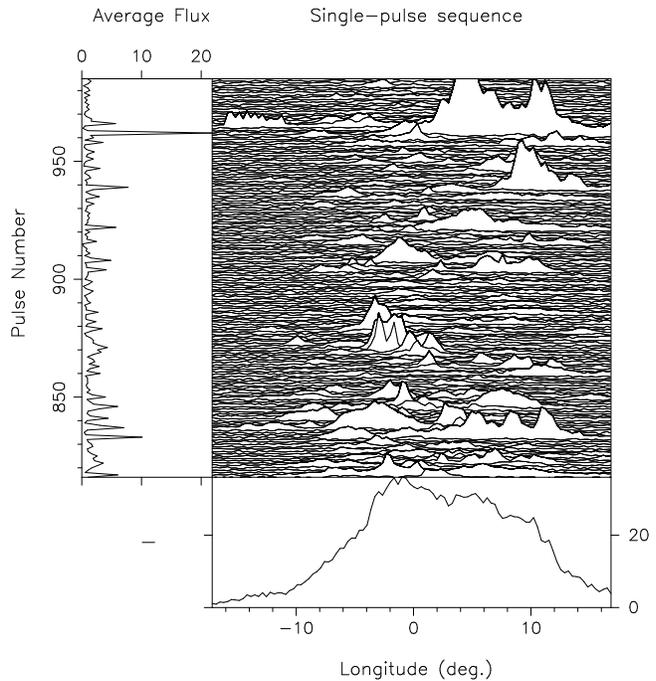}
%\parhang {\bf Figure 12.}---
\caption{Chaotic individual-pulse 
behaviour  during the 1992 ``Q''-mode 
sequence (pulses 817--986). The average 
profile is shown in the bottom panel 
and the pulse-energy variation in the 
left-hand panel.  Note the extremely 
intense subpulses and the enhanced 
emission on the tailing edge of the 
window as compared with the ``B'' mode.
\label{fig:fig12}}
\end{figure}

The 816-pulse sequence thus far considered corresponds 
to the pulsar's ``B'' mode; this is not the case, 
however, for the 170 pulses that follow it.  
Suleymanova \etal\ (1998) studied the attributes 
of these two modal sequences, and they find both 
a sharp boundary between ``B''- and ``Q''-mode 
characteristics at pulse 816/817 and evidence for 
slower variations that both anticipate this 
``mode change'' and follow it.  Here, we will 
first examine the fluctuation properties of this 
short ``Q''-mode sequence, which is displayed 
in Figure~\ref{fig:fig12}---the full sequence, 
average, and pulse-energy variations are given 
in the main, bottom, and left panels, respectively.  
We see again that this ``Q''-mode sequence is 
hardly ``quiescent''; while being weaker overall, 
some individual subpulses are much brighter than 
any observed in the ``B'' mode.   

\begin{figure}
\epsfig{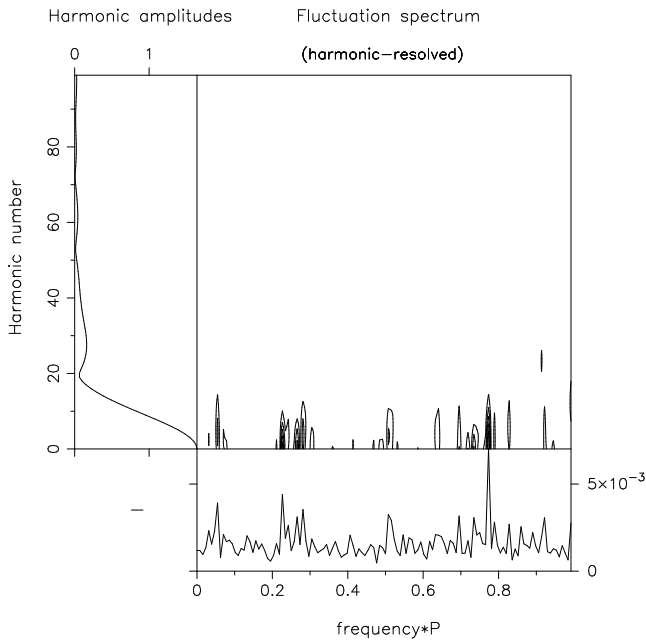}
%\parhang {\bf Figure 13.}---
\caption{Harmonic-resolved spectrum 
for the 128-pulse ``Q''-mode sequence as 
in Fig.~\ref{fig:fig12}.  The intensities 
have been smoothed to reduce the dominance 
of the ultra-strong subpulses.  The spectrum 
clearly indicates that the likely true 
modulation frequency is about 0.77 c/$P_1$, 
which would alias to 0.23 c/$P_1$ in the 
longitude-resolved spectra.  Contour levels 
are about 0.015 mJy$^2$.  
\label{fig:fig13}}
\end{figure}

The intensities of this 128-pulse sub-sequence 
have been smoothed to reduce the effect of the 
ultra-strong pulses.  Its profile is noticeably 
broader, particularly on the trailing edge, 
and its LRF spectra show prominent modulation 
at an apparent frequency of about 0.23 c/$P_1$.  
The HRF spectrum shown in Figure~\ref{fig:fig13} 
clearly indicates that the actual modulation 
frequency is likely to be about 0.77 c/$P_1$, 
which would have a first-order alias at 0.23  
c/$P_1$ in the HRF spectra.  

\begin{figure}
\epsfig{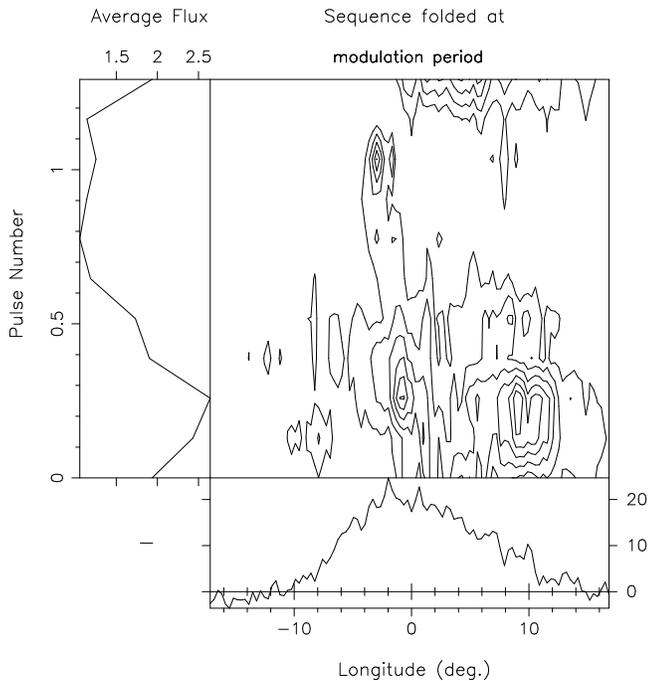}
%\parhang {\bf Figure 14.}---
\caption{Folded sequence corresponding 
to the 128-pulse ``Q''-mode subsequence 
in Fig.~\ref{fig:fig12}.  The data 
have been averaged over 10 bins spanning 
the amplitude-modulation feature at 1.293 
$P_1$/c.  The ``base'' profile (bottom 
panel) has been subtracted from the folded 
sequence displayed (main panel).  The 
modulation is deep and confined to a 
-2$\deg$ to +10$\deg$ longitude range.  
Contours are at 13 mJy intervals.  
\label{fig:fig14}}
\end{figure}

Note that the harmonic amplitudes peak at very
low numbers---about 1 or 2.  If these sidebands 
are interpreted as a phase modulation, then 
the corresponding $P_2$ would definitely be 
in excess of $180\deg$---that is, it would 
be larger than the pulse window and comparable 
to the rotation period.  Therefore, it cannot 
be a phase modulation, so we should attribute 
this feature to amplitude modulation and 
interpret the asymmetry in the sidebands as 
result of a somewhat faster modulation (with 
a frequency of 0.77 c/$P_1$ or 1.77 (2--0.23) 
c/$P_1$).  We will assume the frequency is 
0.77 c/$P_1$ and attempt to verify whether 
this is correct.  Figure~\ref{fig:fig13} shows 
the sequence as folded at this interval, with 
the ``base'' profile (bottom panel) first  
subtracted from the main panel display.  The 
modulation is deep and confined to a -2$\deg$ 
to $10\deg$ longitude interval.  We have 
examined the two halves of this sequence see 
if the behaviour is similar, finding that the 
modulation over the wider part of the pulse is 
common to both subsets, while the feature at 
about $-3\deg$ longitude is present only in 
the first half and is probably due to an odd 
strong pulse.  This amplitude modulation 
feature manifests itself in the pulse-energy 
variations, where we see a tendency for every 
fourth pulse to be more intense than the average 
(see the left panel of Fig.~\ref{fig:fig12}).  

Apart from this prominent feature, a weak 
modulation is also visible near (about 5\% 
less than) the primary ``B''-mode feature 
discussed above, suggesting that the 
``B''-mode modulation is not entirely absent.  
A striking difference between the ``B''- and 
``Q''-mode behaviour is the longitude range 
over which the periodic modulation dominates.  
Later longitudes are weakly modulated in the 
``B'' mode, whereas just the opposite occurs 
here.  Moreover, in the one case we see a 
virtually pure phase modulation, while here  
we have a strong amplitude modulation.  Such 
marked modulation differences would be more 
understandable if, for example, the ``Q'' 
mode were associated with more central 
longitudes than those of the ``B''-mode, and, 
while there is some indication that this 
could be the case, an error in the central 
longitude could produce much the same effect.  
Given all these differences as well as the 
polarisation differences, it is surprising 
that both modes appear to have a nearly 
identical ``base'' profile---where the 
``base'' is defined as that power which 
does not participate in the dominant 
systematic/periodic fluctuations.  

The abrupt change in fluctuation features at 
the ``B''- to ``Q''-mode transition {\it appears} 
to begin with pulse 817, which is both broad 
and very intense. However, the gradual decrease 
in ``B''-mode intensity that precedes the onset 
of the ``Q'' mode (see Suleymanova \etal\ 1998) 
is unlikely to be a coincidence.  Hence, we 
have taken a closer look at the behaviour of 
the sequence just before and after the mode 
change to see if the ostensibly sudden 
transition could have been {\it expected}, 
or whether it is actually as sudden as it 
appears.  We therefore examined the sequence 
prior to the ``Q''-mode onset to see if the 
0.77 c/$P_1$ feature was present, even at a 
low level---and as far as we can determine, it 
is not. 

\begin{figure}
\epsfig{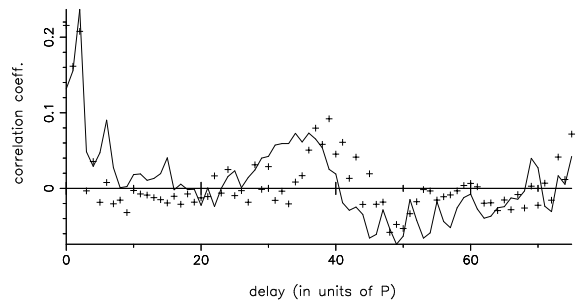}
%\parhang {\bf Figure 15.}---
\caption{(a) Autocorrelation function 
(ACF) of the 128-pulse ``Q''-mode 
subsequence shown of fig.~\ref{fig:fig12}.  
The ACF shown here has been smoothened by 
a 5-lag window to reduce the corrugation 
contributed by the prominent periodicity 
discussed above. (b) Results of a similar 
analysis of ``B''-mode pulses 129--384 
for comparison.  Note the peaks at a lag 
of about 40 periods in both the cases. 
\label{fig:fig15}}
\end{figure}

In an effort to better characterize this 
modal behaviour, we asked whether the 
cartographic-transform images spanning the 
``B''- to ``Q''-mode transition might 
provide some useful insight.  An immediate 
question to assess, however, was whether or 
not the parameters defining the transform 
for the ``B''-mode sequence would be 
appropriate for the ``Q'' mode.  Clearly, 
the geometrical parameters could not change, 
but we had no {\it a priori} means of knowing 
whether the subbeam rotation rate would 
remain fixed.  While we were able to determine 
the round-trip circulation time for the 
``B''-mode regime, the ``Q''-mode sequence 
exhibits no corresponding higher-order 
modulation on which to base an independent 
determination---as would be the case, for 
instance, if the subbeam pattern was not 
stable over a full rotation.  In order to 
see whether the ``Q''-mode pattern is stable 
over at least one circulation time, and, if 
so, to estimate this interval, we examined 
the autocorrelation function (hereafter, 
``ACF'') of the pulse-energy sequence.  The 
``Q''-mode ACF (solid curve) shown in 
Figure~\ref{fig:fig15} has been smoothed by a 
5-sample window to reduce the corrugation of 
the 0.77 c/$P_1$ modulation.  For comparison, 
the dashed curve gives a similar ACF for the 
``B''-mode pulses 129--384.  The peaks at 
lags of about 35--40 periods for the 
``B''-mode sequence are expected; however, 
it is remarkable that they appear to persist 
in the ``Q''-mode sequence as well.  Indeed, 
that both ACFs exhibit correlation at some 
35--40 periods provides very strong evidence 
that the circulation time of the subbeam 
pattern has remained essentially unchanged 
throughout the transition to the``Q'' 
mode---despite the dramatic change in 
modulation properties.  

%Having confirmed that the same transform
%parameters are appropriate for both the 
%``B'' and ``Q'' modes, despite all of the 
%other changes outlined above, we can use it 
%to study the PS.  To begin with, we have 
%applied the transform and its inverse to the 
%``Q''-mode sequence to see if we could 
%reproduce the fluctuation properties 
%satisfactorily.  We were able to do so 
%without any significant change to the transform.  
%However, some weak periodicities get amplified 
%on the closure path, primarily due to the 
%implicit symmetry in the inverse transform.  
%The closure reproduces the spectrum quite 
%well when a circulation time of 20 $P_3$ 
%(= 39.2 $P_1$, calculated from the weak 
%feature at about 0.51 c/$P_1$ in 
%fig.~\ref{fig:fig13}) is used---amounting 
%to a larger interval by some 5\% compared 
%with the 37.3 $P_1$ ``B''-mode period.  
%!!Desh, are the above numbers correct??

Further, we applied the cartographic 
transform to a large number of short 
sequences, each with a large fractional 
overlap.  Viewing the resulting images, 
in a slow ``movie''-like fashion, allows 
us to ``see'' the evolution of the subbeam 
pattern with time.  Since each of the 
images uses a relatively small number of 
pulses, the sampling is thus somewhat 
sparse on the periphery of the polar cap.  
Thus, we have smoothed the image suitably 
to reduce this patchiness, ensuring that 
the basic details are not lost.  In the 
early part of the ``B''-mode sequence, 
most of the 20 subbeams are bright and 
nearly uniformly spaced.  With time, the 
subbeams wane and wax in intensity on 
a scale of several circulation times, 
and overall they gradually decline in 
intensity.  Some either bifurcate or 
partially merge with their neighbors, 
but soon reestablish their characteristic, 
regularly spaced, vigesimal configuration.  
Toward the end of the``B''-mode sequence, 
most of subbeams are weak, with only one 
or two retaining their brightness.  

At the onset of the ``Q'' mode, most of the 
subbeams are weak but still distinguishable; 
however, first one and then two are 
exceptionally strong, with their intense 
discharges filling the area around them 
and spilling into larger radii or heights.  
Their lifetimes appear shorter than $P_1$, 
not allowing adequate sampling, thus 
resulting in their ``streaky'' character.  
These dramatic ``Q''-mode events do not 
entirely displace the old (``B''-mode) 
subbeams, some of which crowd together, 
leaving space in the ring.  In the  process, 
more subbeams appear, whose configuration 
is necessarily more compact, which leads 
to faster fluctuations.  Groups of closely 
spaced subbeams are evident, with spacings 
which correspond to the observed 0.77 
c/$P_1$ feature.  It would appear that 
large intervals of weak activity prompt the 
occurrence of the intense discharges.  It may 
then be that eventually the closely spaced 
subbeams evolve so as to have just enough 
space for 20---thus reestablishing the 
conditions for the bright and stable 
configuration that characterizes the 
``B''-mode sequences.

\section*{IX. Polarisation of the Subbeams}

Both our analyses above and those of Suleymanova 
\etal\ (1998) raise fundamental questions about 
subbeam polarisation:  a) How is the average 
profile comprised of primary-, secondary-polarisation 
mode (hereafter, PPM and SPM), and perhaps 
unpolarised (hereafter, UP) power, and what is 
the relation between the linear and circular 
polarisation (hereafter, LP and CP) in the 
former?  b) How can we account for the longitude 
offset between the PPM and SPM partial average 
profiles?  c) Do the two modal profiles have 
distinct PA sweep rates, and how do these 
combine in the total ``B''-mode average profile?  
d) How does the depolarisation of the star's 
radiation occur, and can we understand it as the 
incoherent mixing of two fully (but orthogonally) 
polarised basis modes?  e) How does the polarisation 
of the rotating beam system differ from that of 
the constant ``base'' profile?  And, finally, 
f) what do the polarisation ``maps'' of the 
subbeams tell us about the conditions of their 
emission?  

In order to explore these questions, we have 
developed several ``mode separation'' techniques.  
Heretofore, such methods have been used to 
segregate a pulse sequence, sample by sample, 
into a pair of modal partial profiles (and 
sometimes a residual profile) [Cordes \etal\ 
1978; Gil \etal\ 1981; Rankin 1988; Rankin 
\& Rathnasree 1995, 1997],\footnote{McKinnon 
\& Stinebring (1998) have carried out a very 
interesting analysis, which also shows that 
incoherent mode-mixing can account for the 
depolarisation in B2020+28.} but for our 
present purposes we require techniques for 
separating the original sequence into a pair 
of polarised modal {\it sequences} (and perhaps 
an unpolarised sequence).  The first method, 
which we have called ``modal repolarisation'' 
assumes that all the observed linear and 
circular depolarisation results from incoherent 
orthogonal-mode mixing.  Under such an assumption, 
the linear and circular depolarisation occurs 
in strict proportion, as (apart from an overall 
angle) the linear can also be regarded as simply 
positive or negative (indeed, as it is represented 
on the Poincar\`{e} sphere).  The object is then 
to restore the observed (depolarised) sequence 
to a hypothetical pair of fully polarised modal 
sequences, preserving both the total power and 
the ellipticity of the original sequence.

\begin{figure*}
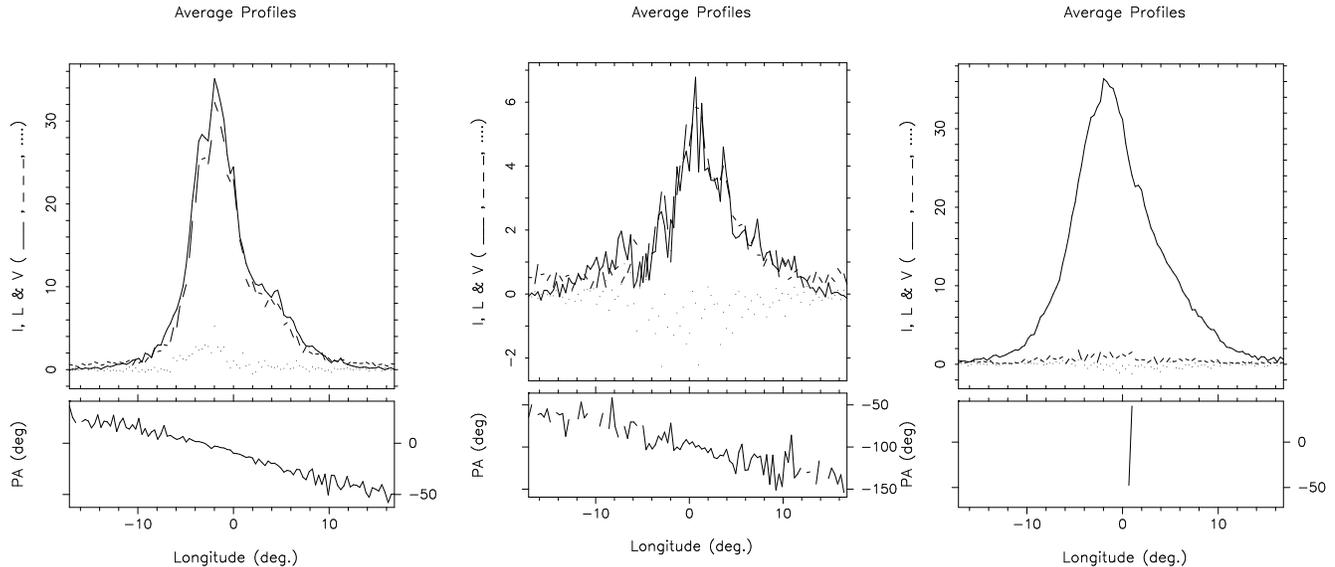

\begin{center}
\begin{tabular}{@{}lr@{}lr@{}}
{\mbox{\epsfig{file=PQmnshort_fig16p.ps,height=5.7cm,angle=-90}}}&
{\mbox{\epsfig{file=PQmnshort_fig16s.ps,height=5.7cm,angle=-90}}}&
{\mbox{\epsfig{file=PQmnshort_fig16u.ps,height=5.7cm,angle=-90}}}\\
%\parhang {\bf Figure 16.}---
\end{tabular}
\end{center}
\caption{Polarised partial-average
profiles corresponding to the PPM, SPM
and UP partial sequences, respectively.
Note that the PPM and SPM profiles are
both highly polarised and elliptically
orthogonal.  By contrast, the UP profile
exhibits a negligible level of polarisation.  
Note also that the PPM and UP sequence 
averages are comparable,both in power 
and in mean (scale)intensity, whereas 
the SPM average is about 5 times weaker.  
The scales are in mJy.
\label{fig:fig16}}
\end{figure*}

In practice these methods proceed by 
examining the total intensity $i$ and total 
linear polarisation $l$ of the sequence, 
sample by sample, and then determining 
whether they respectively fall above or 
below a threshold $t\sigma_N$, where $t$ is 
a chosen value (usually taken to be 2.0) 
and $\sigma_N$ is the estimated on-pulse 
{\it rms} noise level.  If both the quantities 
are above the threshold, then the observed 
PA is checked to see whether it falls 
within $\pm{\pi/2}$ of the model PA.  The 
detailed maths are given in the Appendix.  
We have tested this ``modal repolarisation'' 
method on pulsars for which its motivating 
assumption is best exemplified---B1929+10 
being the canonical such pulsar (Rankin 
\& Rathnasree 1997)---and it functions 
remarkably well both at modest S/N ratios 
and when virtually all of the total power 
is in one of the modes.  Of course, we 
make no claim that these methods are 
{\it precisely} correct in a formal 
statistical sense, but they represent a 
convenient and useful technique for our 
purposes.  

The sequences of some pulsars do not 
divide neatly into two fully polarised 
modal sequences, and 0943+10's provide 
one such instance.  It appears that a 
portion of the polarised and unpolarized 
power in its sequences has a different 
origin, and so we have also found it 
useful to divide the original sequence 
three ways, into two fully polarised 
modal (PPM and SPM) sequences and a 
third unpolarised (UP) sequence.  The 
operation of this ``polarisation 
segregation'' method is quite similar 
to the two-sequence ``repolarisation'' 
method above, and we both discuss its 
details in the Appendix and immediately 
apply it in our analysis and discussion 
below.  

The three panels of Figure~\ref{fig:fig16} 
give the respective ``B''-mode average 
profiles corresponding to the PPM, SPM 
and UP sequences.  The virtually complete, 
elliptically orthogonal polarisation of the 
modal averages---and the negligible levels 
of polarisation in the latter 
average---demonstrate the efficacy of 
our segregation technique.  Note that 
the PPM is about five times greater 
than the SPM, both in power and in peak 
intensity, as the effective widths of 
their respective profiles are about the 
same ($6.7\deg$ vs. $7.1\deg$).  The 
PPM profile leads the SPM one by some 
3$\deg$, and they are slightly left- 
(LHC) and right-circularly (RHC) 
polarised, respectively.  Less than 
half (45\%) of the total ``B''-mode 
power is polarised, but the greater UP 
profile width (~9.7$\deg$) makes the 
modal and UP contributions to the total 
intensity roughly equal at most longitudes.  

The PA traverses of the PPM and SPM in 
Fig.~\ref{fig:fig16} are, overall, quite 
linear, but the curves carry little 
information in the wings of the profiles, 
where the linear power is low and thus 
ever more influenced by the choice of 
the model PA traverse.  Suleymanova 
\etal\ found a very significant difference 
in the PA rates of the ``B'' and ``Q''-mode 
profiles, and, given that the former is 
PPM and the latter SPM dominated, there 
is a case to be made.  We find it hard to 
understand, though, how the PPM--SPM 
difference could be anywhere as large 
as the 50\% (-2.4$\deg/\deg$ {\it vs.} 
-3.6$\deg/\deg$) ``B'' mode--``Q'' mode 
difference that they find.  We have 
examined PA-frequency maps and the 
modal partial profiles above and conclude 
that the PPM {\it might} be as flat as 
-2.4$\deg/\deg$ and the SPM as steep as 
-3.0$\deg/\deg$, but these seem to be 
outside limits.  Even these small 
differences in modal PA rate, however, 
combined with the longitude offset 
between their power centres, is probably 
sufficient to explain the peculiar form 
of the overall PA traverse, which is 
shallow at negative longitudes and then 
steepens significantly at positive ones.  
Note that the rate difference implies that 
the angle between the two modes will 
be significantly less than $90\deg$ at 
~+10$\deg$ longitude, just where the 
overall traverse is steepest.    

\begin{figure}
\epsfig{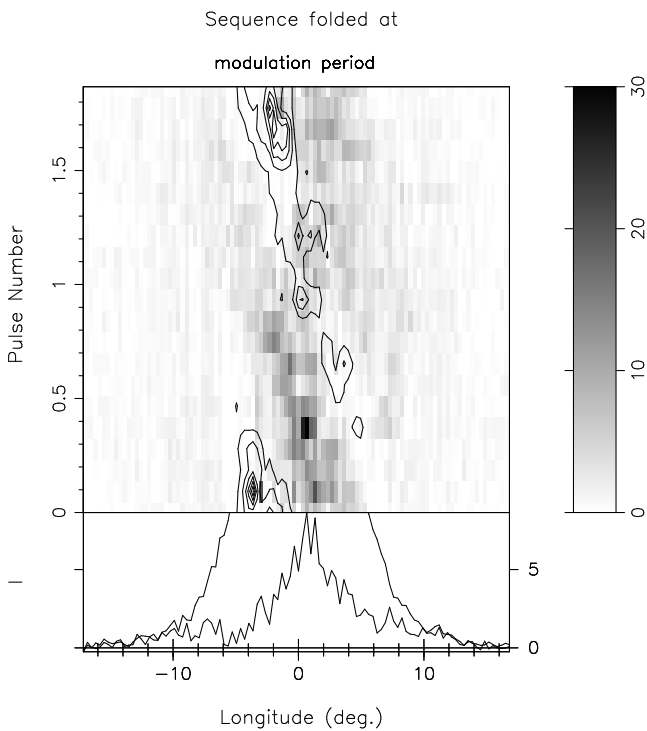}
%\parhang {\bf Figure 17.}---
\caption{Block-averaged pulse sequences 
computed from the respective PPM and SPM 
partial sequences as in Fig.~\ref{fig:fig6}---the 
former shown as a set of contours and 
the latter as a grey-scale plot.  Note 
that both PPM and SPM ``drift'' along 
parallel ``tracks'', but at somewhat 
different phases within the overall 
1.87-period phase modulation cycle.
\label{fig:fig17}}
\end{figure}

Turning now to the polarisation of the 
subpulse ``drift'', the PPM, SPM and UP 
sequences provide ideal instruments 
for identifying its characteristics.  
First, each of these sequences can be 
folded at the basic 1.87-$P_1$ 
phase-modulation cycle ($P_3$) as we 
did in Fig.~\ref{fig:fig6} above.  
Figure~\ref{fig:fig17} then depicts 
the behaviour of the PPM- and SPM-associated 
modulation---the former is given as a 
set of contours and the latter as a 
grey-scale.  Their relationship is 
quite striking.  We know the slopes of 
the ``drift'' tracks quite accurately; 
clearly, on average, the subpulses move 
through -10.5$\deg$ longitude ($P_2$) 
in the 1.867-$P_1$ cycle, or some 
5.62$\deg/P_1$.  

We see in Fig.~\ref{fig:fig17} that 
the PPM and SPM tracks are parallel as 
expected, though we could not have 
anticipated that the latter would fall 
almost exactly in between the former, 
so that PPM and SPM subpulse ``tracks'' 
will always be found at a longitude 
interval of $P_2/2$ or some 5.3$\deg$.  
The overall behaviour is a little more 
complex, however, because there is a 
significant phase difference between 
the PPM and SPM maxima within the phase 
modulation cycle.  The SPM maximum not 
only lags the PPM one by about 0.25 
$P_1$, but the respective intensity 
variations over the cycle appear to be 
almost reflections of each other---that 
is, the PPM exhibits a long slow increase 
and then an abrupt fall, whereas the SPM 
cycle appears to begin with a steep 
increase which then tails off gradually.  

\begin{figure*}
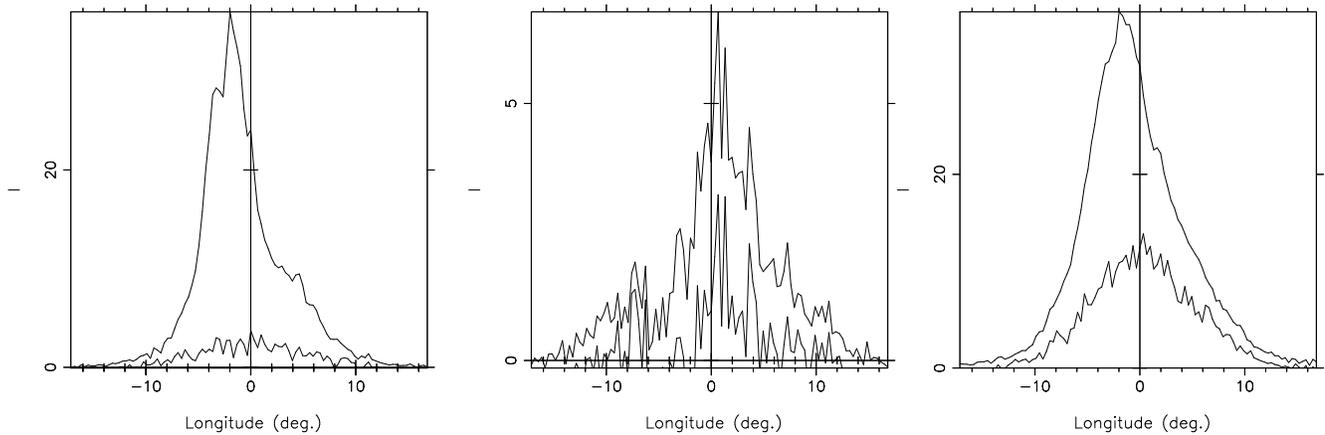

\begin{center}
\begin{tabular}{@{}lr@{}lr@{}}
{\mbox{\epsfig{file=PQmnshort_fig18p.ps,height=5.7cm,angle=-90}}}&
{\mbox{\epsfig{file=PQmnshort_fig18s.ps,height=5.7cm,angle=-90}}}&
{\mbox{\epsfig{file=PQmnshort_fig18u.ps,height=5.7cm,angle=-90}}}\\
%\parhang {\bf Figure 18.}---
\end{tabular}
\end{center}
\caption{Partial-average profiles for 
the PPM, SPM and UP partial sequences 
showing the relative contributions of 
the aperiodically-fluctuating ``base'' 
within each partial sequence.  Note 
that the ``base'' comprises a small 
part of both the PPM and SPM power, 
but constitutes nearly half of the UP 
power.  Note also that it is quite 
symmetrical about 0$\deg$ longitude.  
The ``base'' in each case reflects 
both that power which is steady from 
pulse to pulse as well as any 
which is fluctuating at a different 
rate than the phase-modulation rate. 
\label{fig:fig18}}
\end{figure*}

In order to examine this behaviour in 
more detail, Figure~\ref{fig:fig18} 
gives the result of folding each of 
the three partial sequences at the 
1.867-$P_1$ phase-modulation interval.  
Here we give only the sequence averages 
(full curves) and the relative amplitude 
of the ``base'' profiles inside them.  
This ``base'' has often been subtracted 
in displaying the results of our 
analyses and represents that power which 
is either constant from pulse to pulse 
or fluctuating at frequencies other than 
the phase-modulation rate.  One can 
readily see here that most of the power 
associated with the PPM and SPM is 
fluctuating with the basic 1.867-$P_1$ 
cycle---and therefore associated both 
with the ``drifting'' subpulses and the 
rotating-subbeam system.  Much of the 
UP power is also fluctuating and can 
be seen to largely follow the PPM 
``track'' as expected; though nearly 
half of this power is {\it not} 
fluctuating in this way, and thus 
must have a different origin.  It is 
the sum of these three contributions 
to the ``unfluctuating'' power that 
we encountered above as the triangular 
``base'' profile in Fig.~\ref{fig:fig6}---and 
there questioned whether this power, 
on account of its about 11$\deg$ width, 
could be associated with the weak tail 
of a core component.  

We have also carried out a similar 
analysis using the two partial 
sequences generated by the 
``repolarisation'' technique, and 
the result is telling.  Of course, 
any depolarisation within a sequence 
can be interpreted as orthogonal mode 
mixing and the characteristics of the 
original pair of fully polarised 
sequences then estimated.  For 0943+10, 
however, the large fraction of the 
total power that we saw above in the 
UP sequence is divided equally between 
the modal sequences in the two-way split 
and then (because its fluctuations are 
correlated with those of the PPM) 
dominates the weaker SPM fluctuating 
power in its sequence.  These 
circumstances both make this latter 
technique less useful in 0943+10's 
case and again suggest that some part 
of the UP power originates from a source 
other than the mixing of two fully 
polarised basis modes.  This line of 
interpretation is consistent with the 
UP being, mostly, a weak signature 
of core emission, but certainly does 
not prove that this is the case.  We 
have also computed a Stokes profile 
corresponding to the ``base'' emission, 
and it will come as no surprise that 
this profile shows only slight linear 
polarisation with a PA following the 
SPM traverse and almost no circular 
polarisation.  

\begin{figure}
\epsfig{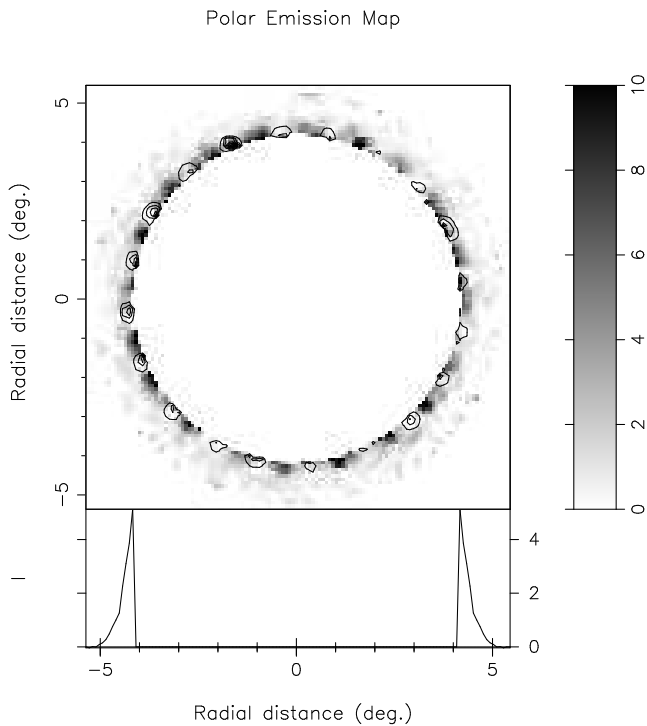}
%\parhang {\bf Figure 19.}---
\caption{Images of the accessible 
emission zone as in Fig.~\ref{fig:fig10}. 
The PPM is shown as a set of contours 
and the SPM as levels of grey.  The 
scales of the two modes are not equal, 
the weaker SPM having been enhanced so 
as to be approximately as prominent as 
the PPM. Note that the SPM power falls 
between the PPM subbeams.  Recalling 
that the subbeam pattern rotates ccw, 
and that the star's rotation sweeps 
the sightline through the top of the 
subbeam system from left to right, we 
see that the SPM emission generally 
lags the PPM beams by just less than 
half the interval between them (see 
text).  
\label{fig:fig19}}
\end{figure}

Of course, each of these sequences can 
also be mapped onto the polar cap using 
the cartographic transform, and the 
relation between the PPM and SPM beams 
is shown in Figure~\ref{fig:fig19}.  
Here, the PPM emission configuration is 
shown as a set of contours and that of 
the SPM as a grey-scale map.  The 
respective intensity scales are, of 
course, not equal; indeed, the SPM 
scale has been enhanced by a factor of 
about 5 for clarity.  

Note, first of all, that the SPM 
emission also forms a set of subbeams 
along nearly the same circle as that 
of the PPM.  Indeed, we know that all 
these subbeams rotate ccw about the 
magnetic axis at the centre of the 
diagram and that the star's overall 
cw motion about its rotation axis (at 
the top of the diagram) causes the 
sightline path to traverse from right 
to left through the beams, just 
poleward of the magnetic axis.  The 
different longitude centres and 
modulation phase of the SPM and PPM 
power together determine the relative 
(azimuth) orientation of their 
respective subbeam systems.  On 
average, the SPM subbeams lag their 
preceding PPM neighbors by an 
interval which is a little less than 
half the mean subbeam spacing.  It is 
just this lag which is reflected in 
the delay of the SPM profile power 
centre relative to that of the PPM in 
Fig.~\ref{fig:fig16}.  Indeed, we 
see there that the SPM power centre 
follows the PPM one by 3--4$\deg$ out 
of the 10.5$\deg$ spacing between 
subpulse ``drift'' bands ($P_2$).  
Note also that both the SPM and PPM 
beams vary markedly in mean intensity, 
and there is some correlation between 
the strength of an SPM beam and that 
of the PPM beams adjacent to it.  Again, 
on average, the two beam systems 
both seem truncated at the absolute 
limit imposed by the sightline 
traverse---and to about the same 
extent.  

\begin{figure}
\epsfig{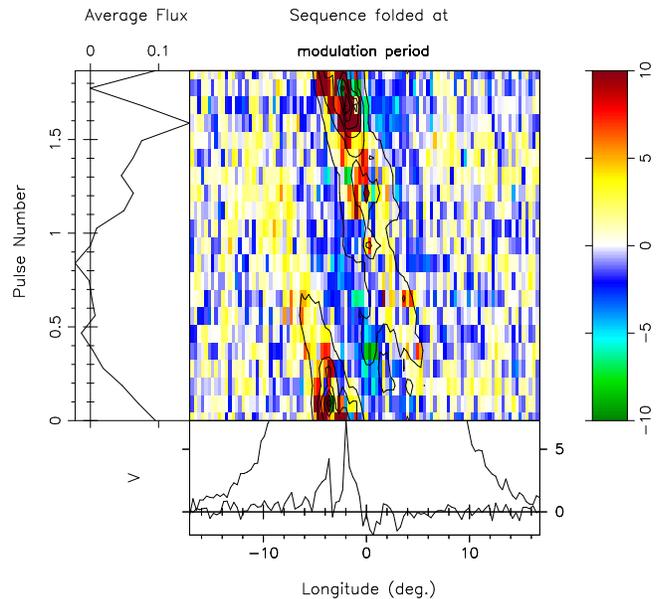}
%\parhang {\bf Figure 20.}---
\caption{Block-averaged pulse 
sequences computed from Stokes 
parameters $I$ and $V$, respectively, 
as in Fig. ~\ref{fig:fig6}---the 
former shown as a set of contours 
and the latter as a colour plot.  
Note that the PPM-dominated $I$ 
``drifts'' along a ``track'' that 
has significant LHC polarisation, 
while RHC lies along the adjacent 
SPM ``track''.  The circular 
polarisation can be quite strong 
at certain positions along these 
``tracks'', but overall remains at 
a low level. 
\label{fig:fig20}}
\end{figure}

\begin{figure}
\epsfig{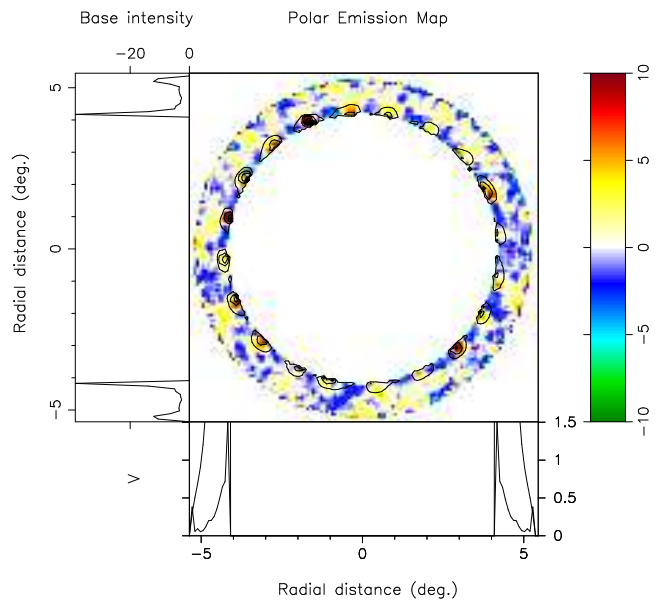}
%\parhang {\bf Figure 21.}---
\caption{Images of the accessible
emission zone as in Fig.~\ref{fig:fig10}.
Stokes $I$ is shown as a set of contours 
and Stokes $V$ as a colour plot.  Note 
that the PPM-dominated $I$ subbeams are 
LHC polarised, whereas the SPM-associated 
region between them is RHC polarised.  
Each of these circularly polarised regions 
has a tendency to extend to ``outside'' 
regions, which may be indicative of higher 
altitudes in the emission region. 
\label{fig:fig21}}
\end{figure}

Finally, we can locate the circular 
polarisation within the ``drift'' 
bands and within the subbeam system 
in a similar fashion.  Figure
~\ref{fig:fig20} again shows the 
result of folding the sequence at 
the 1.87-$P_1$ modulation-cycle 
interval, but here we see the total 
power (Stokes $I$) plotted as a 
contour map and the circular 
polarisation (Stokes $V$) given as 
a colour scale.  It is immediately 
clear that positive (LHC) circular 
polarisation is associated with the 
total power---which in turn is 
dominated by the PPM emission; 
whereas bright negative (RHC) 
circular polarisation is seen along 
the parallel ``track'' corresponding 
to the SPM.  Note that the fractional 
circular can reach high levels 
at certain points in the diagram 
for both modes, but that its average 
level remains low.  Of course, the 
implied beam configuration of this 
circular polarisation can also be 
viewed using the cartographic 
transformation, and Figure
~\ref{fig:fig21} gives this 
result.  Again, the total power is 
shown as a set of contours and the 
circular as a colour scale.  Here 
also, we see LHC polarisation 
associated with the PPM beams 
as well as the region ``outside'' 
them---that is, further from the 
magnetic axis or perhaps at greater 
height along the same fields lines.  
Similarly, the RHC is found just 
between the PPM beams near the 
positions of the SPM beams in Fig.
~\ref{fig:fig19} and also 
``outside'' them.  

Overall, the results of our analysis in 
this section are quite clear.  The rotating 
subbeam system that we encountered in 
earlier sections is highly polarised in an 
elliptically orthogonal manner:  the 
prominent 20-fold subbeam system is 
dominated by PPM emission, which also 
exhibits systematic LHC polarisation.  
Then, interleaved with this PPM beam 
system is a similar, weaker SPM beam 
system which we find to have significant 
RHC polarisation.  These circumstances 
appear to suggest a set of very specific 
conditions within the emission region, 
and we will come back to a discussion of 
their implications later.

\section*{X. The 1972 January 430-MHz Sequence}

In this section, we look at a sequence 
which was recorded at Arecibo in early 
1972 (Backer \etal\ 1975) in order to 
explore whether the conditions 
encountered in the 1992 observations 
are typical or unusual.  We have 
assessed the quality carefully and 
have selected a subset comprised of 
pulses 201--968 (using only the 
right-circularly polarised channel as 
%Is this right???
the other was corrupted by 60-Hz 
interference) from this old 1000-pulse 
polarimetric observation.  Here, both 
the single-pulse S/N and resolution 
are poor relative to the newer 
sequence---so we used a 3-sample 
smoothing to improve the S/N (at 
the cost of poorer resolution) but 
retained the original sampling.  
Here, too, we find the gradual, 
monotonic decline in the pulse 
energy with time, much like what 
was seen in the newer sequence. The 
average profile, however, has a more 
symmetric appearance, which is 
intrinsic and not a result of the 
applied smoothing.  

\begin{figure}
\epsfig{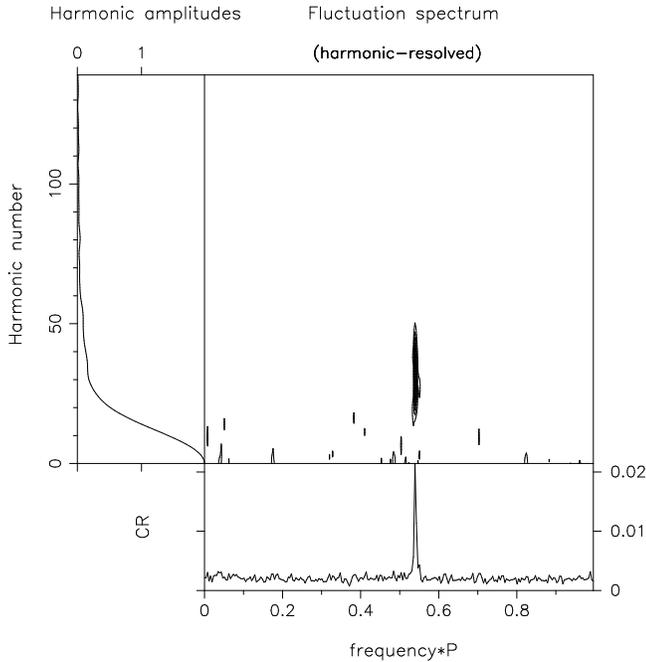}
%\parhang {\bf Figure 22.}---
\caption{Harmonic-resolved fluctuation 
spectrum as in Fig.~\ref{fig:fig4} for the 
1972 430-MHz subsequence.  Note that the 
primary modulation feature is at 0.54 
$c/P_1$---related as before to the {\it aliased} 
0.46-c/$P_1$ feature in the LRF (not shown).  
Contours are at intervals of 28 mJy$^2$ up to 
a maximum of 198 mJy$^2$. 
\label{fig:fig22}}
\end{figure}

In order to examine the fluctuation 
properties of the sequence, we computed 
both the LRF and HRF spectra.  The 
latter is given as Figure~\ref{fig:fig22} 
and its dominant spectral feature at 
$0.5409\pm0.0009$ c/$P_1$ appears in 
the LRF at 0.46 c/$P_1$ (as a first-order 
alias as discussed in \S III).  These 
features correspond to the phase 
modulation associated with the rotating 
subbeams.  The second-harmonic feature 
is not visible in these spectra, but 
is present when the smoothing is not 
performed.  The harmonic amplitudes 
of the fluctuation features in the HRS 
are centred at harmonic numbers 32 and 
around 65, implying a typical subpulse 
separation of about $11\pm0.5\deg$.  
The sequence was also folded at the 
primary modulation period in the manner 
of Fig.~\ref{fig:fig6} and exhibited a 
nearly identical behaviour.  The drift 
band was clearly visible, once the 
``unchanging'' base contribution was 
removed from that folded over the 
modulation cycle.  The modulation period 
($P_3$) used here corresponds to a refined 
value ($(1-1/2.153)^{-1}$ $P_1$)at which 
the modulation pattern shows maximum 
contrast in pulse-energy variation across 
the modulation cycle.  Here there was also 
a slight suggestion that the subbeams cross 
our sightline twice during each complete 
circuit in magnetic azimuth; if so, it 
would imply that their angular distance 
from the magnetic axis is larger than 
$|\beta|$, which is consistent with many 
subaverage profiles  showing a ``conal-double'' 
appearance.  Also, since the two crossings 
through our sightline are generally expected 
to be symmetric with respect to the ``true'' 
longitude origin, they provide a useful 
and independent estimate of the central 
longitude value which we use.  The ``base'' 
profile in this subsequence, though, had a 
larger fractional intensity and a half-power 
width of about $10\deg$, consistent with 
the ``base'' width in the newer observations.  
%!!Desh, the various values in this section 
% and in Table I are not consistent.  37.014 
periods is compatible with a feature at   
% 0.5403 c/P and the 2.153 value above seems 
% to come from the 1992 analysis, not anything 
% here.  Can you tell me how I can patch this up?

We also computed the phase variation of the 
modulation feature (at $f_3$) as a function 
of longitude, and the results were virtually 
identical to Backer \etal's fig. 4 (though we 
have taken care to specify the central 
longitude correctly).  The rate of change of 
the modulation phase with respect to longitude 
gives an independent estimate of $P_2$, which 
we find encouragingly consistent with that 
suggested by the HRF spectrum as well as with 
that found for the 1992 sequence.  Based on 
this similarity, we estimate [using eq. (3)] 
the secondary modulation period to be about 20 
times $P_3$. 

Unfortunately, the relatively noisy character 
of this sequence has frustrated our efforts 
to use the intensity autocorrelation function 
to confirm our estimate for the round-trip 
circulation time.  The modulated intensity here 
is a smaller fraction of the average intensity 
than in the 1992 observation.  Also, there is 
noticeably more emission at longitudes 
corresponding to the profile wings ({\it i.e.} 
at outer radii) where the intensity fluctuations 
are high, but it does not appear to have any 
definite periodicity.  This ``outer'' emission 
is clearly seen in the average profile ``wings'' 
as a pair of weak, roughly symmetrical features.  

\begin{figure}
\epsfig{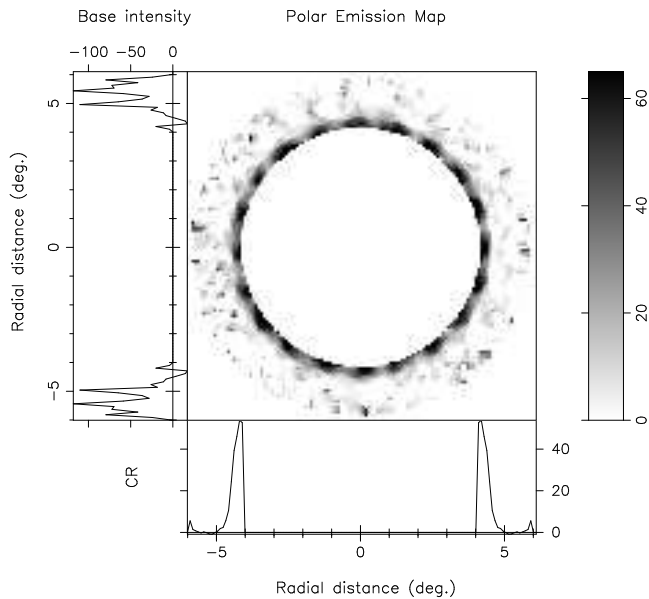}
%\parhang {\bf Figure 23.}---
\caption{Polar emission map obtained via 
the cartographic transform for the 1972 
430-MHz observation in Fig.~\ref{fig:fig10}.  
The map is smoothed with a circular 
8-pixel-area window to reduce the noise in 
its peripheral regions.  Again, only the 
RHC channel was used in the analysis.  
\label{fig:fig23}}
\end{figure}

We have attempted to map the average 
subbeam distribution of this sequence, 
using a circulation time of 37.014 
$P_1$ and the ``inside, inner'' 
geometry discussed earlier.  The 
central longitude estimate was 
verified using a bootstrap procedure 
based on the inverse transform 
``closure''.  Figure~\ref{fig:fig23} 
shows the result after smoothing the 
map with a circular function having 
an area 8 pixels to reduce its noisy 
appearance.  The overall character 
of the distribution is quite similar 
to that of the 1992 ``B''-mode 
sequence; however this map shows 
significant emission at outer 
radii---probably from an outer ring 
of more or less randomly spaced 
subbeams.  This ``outer'' emission 
is different in kind from that of 
the ``Q'' mode at a similar radius, 
basically because its distribution 
is so  ``conal'' but random in 
azimuth.  

To summarize, this ``old'' sequence 
seems to show all the primary  
``B''-mode characteristics, together 
with some weak and random emission 
possibly associated with the same 
sources as the ``Q''-mode emission.  
The secular decrease in pulse energy 
may well be an indication of an 
intense, but sporadic ``Q''-mode 
sequence to follow.  The modulation 
frequency is a little higher than 
that of the recent observation, but 
it is well within the variations 
seen in short sections of the more 
recent sequence.

\section*{XI. The 1990 January 111.5-MHz Sequence}

Now we examine at a sequence observed 
at a longer radio wavelength.  It 
consists in more than 1000 pulses, and 
we have looked at the first 1024 pulses 
in detail.  At this lower frequency the 
profile has two well resolved components 
separated by about 10$\deg$ longitude, 
implying that here the radiation-cone 
radius $\rho$ is larger than the impact 
angle $\beta$.  This is not at all 
surprising if we are observing emission 
from the same subbeams whose outer 
periphery we imaged at 430 MHz.  We would 
expect to see them more fully at lower 
frequency, because of the ever larger 
emission-cone radius.  Indeed, our 
favored geometrical model in Table 
II indicates a conal beam radius $\rho$ 
some 0.26$\deg$ larger than at 430 MHz,
or, said differently, $\beta/\rho$ is 
here 0.96 as opposed to 1.02 at the 
higher frequency.  This is a major 
advantage; and we will then hope to 
``see'' a larger portion of the actual 
radial extent of the subbeams.  

\begin{figure}
\epsfig{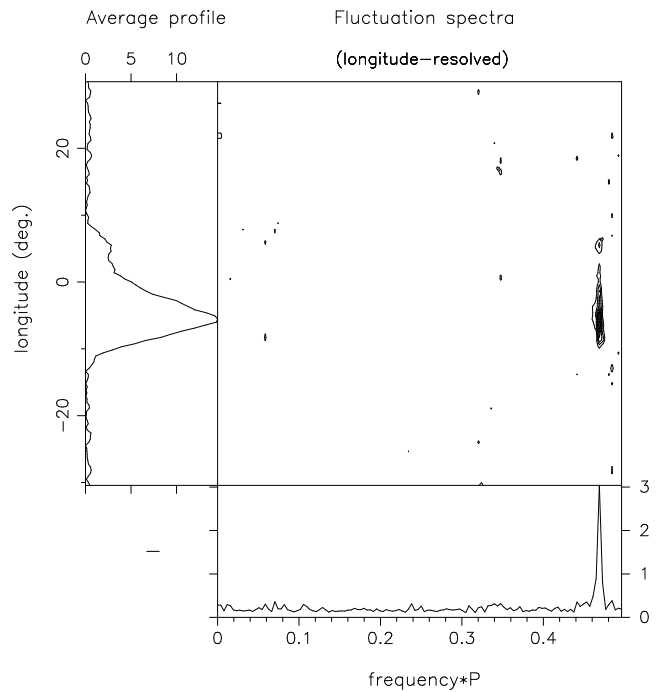}
%\parhang {\bf Figure 24.}---
\caption{Longitude-resolved spectra (using 
256-point ffts) for a subsequence (pulses 
1--350) of the 1990 111.5-MHz observation 
(main panel) and the total spectrum (bottom 
panel) as well as the average profile (left 
panel).  Note the similarity between these 
spectra and those of the 430-MHz observations.  
The intensity units are arbitrary.  
\label{fig:fig24}}
\end{figure}

Let us first examine the fluctuation 
properties of this sequence.  The 
dominant feature, even at this 
frequency, is associated with the 
nearly alternate-pulse modulation 
that we meet yet again. Over the 
1024-pulse sequence, the fluctuation 
frequency varies by perhaps 4\% of 
its mean value, but appears stable 
over intervals of a few hundred 
pulses.  We show, as a typical 
example, detailed spectra for pulse 
numbers 1--350.  Other portions of 
the observation have similar spectra, 
except for small shifts in the 
modulation frequency.  Figure~\ref{fig:fig24} 
shows the LRF spectra, where the 
aliased feature appears at a 
frequency of about 0.46 c/$P_1$. 
The HRF spectrum (not shown) confirms 
that the 0.46-c/$P_1$ feature is 
again the first-order alias of a 
0.53-c/$P_1$ modulation and that it is 
phase modulated.  Its second harmonic 
was found close to 0.07 c/$P_1$.  
The harmonics related to both the 
rotational and the phase modulation 
show a distinctly different envelope 
compared to their equivalents at 430 
MHz.  The difference is a direct 
reflection of the double-resolved 
profile as against the single form at 
430 MHz.  The LRF spectrum confirms 
that the harmonics of the modulation 
feature peak at about 32 c/$P_1$, again 
implying a $P_2$ between 10--$11\deg$.

\begin{figure}
\epsfig{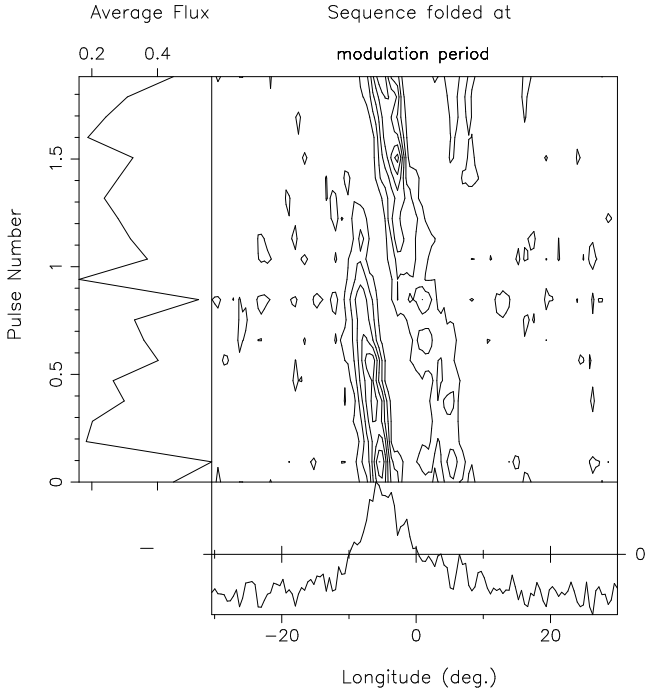}
%\parhang {\bf Figure 25.}---
\caption{Subsequence folded over the  
modulation cycle for the same 1990 111.5-MHz 
observation as in Fig.~\ref{fig:fig24} 
(main panel); the ``base'' profile (bottom 
panel) has been removed.  Note the two 
distinct tracks associated with the two 
components seen in the 111.5-MHz average 
profile.  This should be contrasted with 
the single ``drift track'' seen at 430 MHz 
(see Fig.~\ref{fig:fig8}).
\label{fig:fig25}}
\end{figure}

Next, we look at the result of 
folding the 350-pulse sequence at 
the corresponding modulation period 
(see Figure~\ref{fig:fig25}).  A 
prominent drift band associated with 
the leading component is seen in 
the left half of the main panel.  
Appearing as a faint continuation 
is the ``track'' associated with the 
``drift'' of the weaker trailing  
component.  The ``base'' profile 
subtracted from the data in the main 
panel is given in the bottom panel.  
Note that it here resembles the 
average profile, unlike what we 
observe at 430 MHz.  It is more likely, 
therefore, that this ``base'' reflects 
aperiodically-modulated residual 
emission, stemming possibly from 
variations between the drift-bands of 
different subbeams, rather than 
being the result of another form of 
radiation.    The variation in the 
modulation phase (not shown) is 
remarkably similar to that seen at 
430 MHz, implying about the same value 
of $P_2$ as determined earlier.  The 
conclusions based on the first set of 
350 pulses are seen to be valid 
throughout the entire sequence, apart 
from small differences in the modulation 
frequency.  It may be worth recalling 
that such differences, though on a 
reduced scale, were also evident in 
the recent 430-MHz sequence.    

\begin{figure}
\epsfig{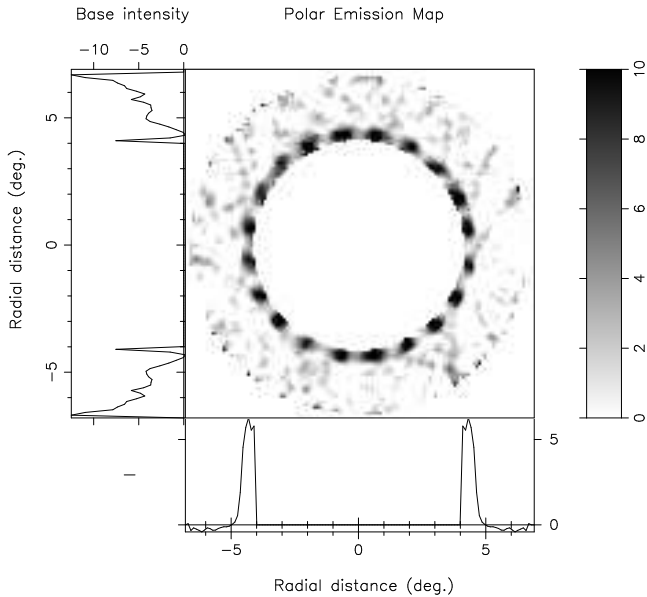}
%\parhang {\bf Figure 26.}---
\caption{Polar map obtained via the 
cartographic transform of the same 1990 
111.5-MHz subsequence as in Fig.~\ref
{fig:fig24} (main panel).  The map has 
been smoothed with a circular 
(8-pixel-area) window to reduce the 
noise in the peripheral, poorly sampled 
areas.  Note the less elongated
appearance of the subbeam features as 
compared to those in the 430-MHz maps, 
implying that the more interior 111-MHz 
sightline samples the beams more fully.  
The intensity scales are arbitrary.  
\label{fig:fig26}}
\end{figure}

We now proceed to perform the 
cartographic transformation.  
Having noted that the modulation 
properties are nearly identical 
to those for the 430-MHz 
``B''-mode sequence, the 
conclusion that the round-trip 
delay is quite close to 20 times 
the mean $P_3$ is unavoidable.  
However, since the differences 
in the $P_3$ values for the 
different sections of this data 
are found to be different by more 
than the available resolution 
in $P_3$ (which is simply the 
inverse of the number of 
pulse-periods in the sequence), 
we should perform the transform 
on suitable sections of the data 
separately, unless we wish to view 
only the average distribution of 
the subbeams.  With the geometry 
already defined, we use the 
different data sections and the 
inverse transform ``closure'' to 
confirm our estimate of the central 
longitude, and then examine the 
subbeam distributions (in simple 
projection onto the polar cap) for 
the different sections separately.  
Figure~\ref{fig:fig26} shows one 
such map, again corresponding to 
the first 350 pulses.  The other 
sections of the data show a similar 
distribution of subbeams, differing 
only in their detailed intensities 
and the locations of individual 
subbeams.  As anticipated, we now 
see a much larger fraction of each 
of the subbeams than was possible 
at 430 MHz.  The radial profile 
(in the bottom panel) shows an 
intensity maximum well outside 
the $\beta$-cutoff.  In fact, 
the fraction of the subbeam that 
is not visible now is quite 
comparable to that visible at 430 
MHz.  By comparing the radial 
profiles at the two frequencies, 
we can now estimate the factor by 
which the radiation-cone size has 
changed---1.03 between 430 and 
111.5 MHz, which is in good agreement 
with the modelling in Table II.

\section*{XII.  Summary and Conclusions}

In this first paper of a series, we develop 
techniques and present detailed studies of the 
drifting subpulses of pulsar B0943+10, based 
on three sets of Arecibo observations.  Pulse 
sequences from this star are remarkable for 
their extraordinary stability and apparent 
complete lack of null pulses.  Future papers 
in this series will feature sequences of several 
different kinds from other low frequency 
observatories as well as overall analysis and 
physical interpretation of their significance.  

Our main results may be summarized as follows: 
\begin{itemize}
\item We have resolved the question of whether 
0943+10's primary spectral feature is aliased and 
whether the secondary feature is its second 
harmonic:  The primary feature at about 0.46 c/$P_1$ 
is the first-order alias of the fundamental 
phase-modulation, which has a true frequency of about 
0.53 c/$P_1$.  The secondary feature is then indeed  
a second harmonic (at about 1.07 c/$P_1$), which is 
seen as a second-order alias at about 0.07 c/$P_1$. 

\item Three new techniques have been developed and 
applied to resolve the above questions: a) A ``harmonic 
resolved fluctuation'' spectrum uses the information 
within the finite width of the pulse to achieve a 
Nyquist frequency of 1 c/$P_1$, showing clearly that 
the primary feature is aliased.  Then, the harmonicity 
of both b) the harmonic amplitudes of the two features 
and and c) their phase rates are used to argue that 
the former appears as a first-order alias and the 
latter as a second-order one.

\item On this basis, we were able to determine that 
$P_3$ is some 1/(0.53 c/$P_1$) or 1.87 $P_1$ and to 
demonstrate this circumstance by folding the entire 
sequence at this interval.  This implies that that 
drift is {\it negative}, or, in the same direction 
as the star's rotation.  

\item Using several different techniques, we were 
able to measure the longitude interval between adjacent 
subpulses $P_2$, as well as the polarisation-angle 
rotation corresponding to this interval $\chi_{P_2}$, 
and to relate these to the azimuthal angle between 
adjacent between subbeams around the magnetic axis 
of the star $\eta$.  The most reliable values of 
$P_2$ come from either the principal harmonic number 
of the phase modulation or the phase rate associated 
with its primary feature, both of which yield values 
of about 10.5$\pm$0.5$\deg$.  Interestingly, we find 
that $|\chi_{P_2}|$, which just less than 30$\deg$, 
should exceed $P_2$ for an inner sightline traverse, 
and {\it vice versa}; and that $\eta$ is approximately 
the difference between these quantities.  Thus $\eta$ 
must be near 18$\deg$ for 0943+10, implying a system 
of some 20 subbeams.  
 
\item A pair of ``sideband'' features are observed in 
the overall fluctuation spectra (and very strongly in 
certain sections).  We identified these features as an 
amplitude modulation on the primary phase modulation.  
Such features can only occur if this tertiary amplitude 
modulation is stable and harmonically commensurate with 
the primary phase modulation.  The separation between 
the ``sidebands'' and the primary feature $\Delta f$ 
is just 1/20 of the 0.54 c/$P_1$ frequency, indicating 
that the tertiary modulation is produced by a repeating 
pattern of just 20 elements.  The overall modulation 
period is then 20 times $P_3$ or just over 37 $P_1$, 
which was dramatically demonstrated by folding the 
entire sequence at this interval.  It is at this point 
that we have conclusive evidence that the aliasing 
question is resolved, because 0.5355/$\Delta f$ is an 
integer within its errors, whereas other possible alias 
frequencies of the primary phase modulation are not.   

\item Pulsar 0943+10's drifting-subpulse pattern can be 
completely understood as resulting from a system of 20 
subbeams rotating around its magnetic axis in an interval 
of 37 rotation periods or about 41 seconds, where some of 
the subbeams are stronger than others and maintain this 
intensity difference for several circulation times.  

\item We have developed a new technique, involving a 
``cartographic'' transform and its inverse, to map 
and study the underlying subbeam pattern.  The forward 
transform merely expresses the observed pulse sequence in 
terms of a pulsar-frame magnetic colatitude and azimuth, 
rotating at the subbeam circulation rate, rather than the 
usual parameters of pulse number and phase (or longitude).  
Techniques are described whereby the forward transform 
can be used to study and display the sequence, whereas 
the inverse transform can be applied in a ``search'' mode 
to determine the geometry or other poorly known parameters 
of the sequence.  

\item All the techniques including the ``cartographic'' 
transforms were applied to three different Arecibo 
observations, two at 430 from 1992 and 1972 and one 
at 111 MHz from 1990, and entirely compatible results 
were obtained for each.  

\item The polarisation characteristics of the most 
recent observations were also carefully investigated.  
Using new methods of segregating the observed sequences 
into two or three subsequences with definite polarisation 
states, we have mapped the pulsar's emission in the 
two polarisation modes and plotted their characteristics.  
The PPM is about five times stronger than the SPM, and 
both closely follow a R\&C behaviour at every longitude 
across the profile\footnote{We see no evidence of the 
anomalous PA rotation within the drifting subpulses as 
has been reported for 0809+74 by Taylor \etal\ (1971) 
and for 2303+30 by Gil (1992). }---though they appear 
to have slightly different PA sweep rates.  Positive 
circular polarisation is associated with the PPM and 
negative with the SPM.  A part of the unpolarised power, 
which drifts, seems to be the result of modal depolarisation, 
but another part seems to have an independent origin. 

\item Polarised maps of the beam configuration show 
clearly that the PPM emission consists of a 20-fold 
system of bright, well confined subbeams, whereas the 
SPM emission is observed as a series of ``beams'' 
interleaved between the PPM system as well as at 
larger radii (or heights).  This and other features 
of the beam configuration suggest that the pulsar 
radiates at various altitudes within a system of 
plasma columns which have their ``feet'' near 
the stellar surface.   
   
\item Our qualitative result that the pulsar's drifting 
subpulse emission is produced by a rotating subbeam system 
as well as our quantitative values for both the circulation 
time and physical subbeam spacing (see also Deshpande \& 
Rankin 1999{\it a}) appear fully compatible with the 
Ruderman \& Sutherland (1975) model, though our analysis is 
completely independent of this model.  

\item The methods of the present analysis should be useful 
for studying the origin of pulse modulation in other pulsars.  
   
\end{itemize}

\noindent {\bf Acknowledgements:} We are grateful to 
V. Radhakrishnan and Rajaram Nityananda for many insightful 
comments and criticisms on our analysis.  We also thank:
Jonathan Arons, Alice Harding, Vinod Krishan and Malvin 
Ruderman for highly informative discussions; R. Ramachandran 
and P. Ramadurai for their generous assistance with computing; 
Svetlana Suleymanova for drawing our attention to this pulsar; 
and N. Rathnasree, Vera Izvekova, Svetlana Suleymanova, 
Kyriaki Xilouris for help with the 1992 observing; and Phil 
Perillat for his Arecibo 40-MHz Correlator software.  We each 
want to acknowledge the hospitality of each other's institutes, 
when much of this work was carried.  This work was supported 
in part by grants from the US NSF (AST 89-17722 \& INT 93-21974).  
Arecibo Observatory is operated by Cornell University under 
contract to the US NSF.

\section*{Appendix.  Polarisation-Modal Analysis Techniques }

Here we give the mathematical details 
of the polarisation-mode separation 
techniques which we used in \S IX.  
As these techniques proceed sample by 
sample, we are always concerned with 
the polarisation state of each individual 
sample, which can be represented by 
Stokes vector $s=(i,q,u,v)$ in relation 
to the standard deviation of all sources 
of noise in any one Stokes parameter
$\sigma$.  The total polarised power 
is then $p=\sqrt{q^2+u^2+v^2}$, and 
the linear power $l=\sqrt{q^2+u^2}$, 
where $\phi={1\over 2}\tan^{-1}(u/q)$ 
is the sample polarisation angle (PA).  
Both techniques require a model PA 
traverse, which is usually taken to 
describe the primary polarisation mode 
(PPM) and which, for the sample in 
question, has a value $\phi_m$.  

Our modal ``repolarisation'' technique 
follows that outlined for profile 
polarisation in Rankin \& Rathnasree 
(1997); however, we here apply it to 
the individual samples of a pulse 
sequence.  To this end we require a 
few more definitions:  The total 
fractional polarisation $f = p/i$, 
so, for the primary mode $l^{\prime}=l/f$ 
and $v^{\prime}=v/f$.  Then, for the 
secondary mode $l^{\prime\prime}=-(l^{\prime}-l)/2$ 
and $v^{\prime\prime}=-(v^{\prime\prime}-v)/2$.  
In that the two modes are fully polarised 
by assumption, 
$i^{\prime}=\sqrt{{l^{\prime}}^2+{v^{\prime}}^2}$ and 
$i^{\prime\prime}=\sqrt{{l^{\prime\prime}}^2+{v^{\prime\prime}}^2}$.  
Finally, $q^{\prime}=l^{\prime}\cos 2\phi$ and 
$u^{\prime}=l^{\prime}\sin 2\phi$, just as 
$q^{\prime\prime}=l^{\prime\prime}\cos 2\phi$ and 
$u^{\prime\prime}= l^{\prime\prime}\sin 2\phi$.  

Insignificant portions of the resulting 
sequences must be noise-like, just as 
are the natural ones.  Therefore, $\hat{n}$ 
here represents a noise source---that is a 
Gaussian-distributed random variable with 
zero mean and a standard deviation equal 
to $\sigma/2$. In generating the two 
repolarised modal sequences, four different 
computations are carried out depending 
upon whether the sample falls above or 
below two different thresholds, one 
pertaining to its total intensity 
relative to the noise level $\sigma$ 
and a second to determine whether it 
is significantly linearly polarised. 
Only when a sample has significant 
power but negligible polarisation is 
the model angle used to reconstruct 
the modes; then we compute the necessary 
linear $l^*=\sqrt{i^2-v^2}$ and so 
$q^*=l^*\cos 2\phi_m$ and $u^*=l^*\sin 2\phi_m$. 
The overall procedure is then as follows:  

\begin{center}
\begin{tabular}{cc}

$l \ge 2\sigma$ and $p \le i$ & $l \ge 2\sigma$ and $p > i$ \\
(partial polarisation)   &  (hyperpolarisation)  \\
$s_{\rm p(s)} = (p^{\prime},q^{\prime},u^{\prime},v^{\prime})$ & $s_{\rm 
p(s)} = (i,q,u,v)$ \\
$s_{\rm s(p)} = 
(p^{\prime\prime},q^{\prime\prime},u^{\prime\prime},v^{\prime\prime}) 
$ & $s_{\rm s(p)} = (\hat{n},\hat{n},\hat{n},\hat{n})$ \\

For $|\phi-\phi_m| < {\pi\over 4}\ (\ge {\pi\over 4})$ & For $|\phi-
\phi_m| < {\pi\over 4}\ (\ge {\pi\over 4})$ \\ \\

$l < 2\sigma$ and $i > 2\sigma$ & $l < 2\sigma$ and $i < 2\sigma$ \\
(depolarisation)  &  (noise)  \\
$s_p = (i/2,q^*/2,u^*/2,v/2)$ & $s_p = (i/2,q/2,u/2,v/2)$ \\
$s_s = (i/2,-q^*/2,-u^*/2,v/2)$ & $s_s = (i/2,q/2,u/2,v/2)$ \\
 &  \\
%& \hspace{1.0in} ...(A.1) \\
\end{tabular}
\end{center}

The three-sequence polarisation 
segregation method is slightly 
simpler to carry out.  The same 
thresholds are used, $\hat{n}$ is 
defined as $\sigma/3$, and the 
overall procedure is as follows: 

\begin{center}
\begin{tabular}{cc}

$l \ge 2\sigma$ and $p \le i$ & $l \ge 2\sigma$ and $p > i$ \\
(partial polarisation)   &  (hyperpolarisation)  \\
$s_{\rm p(s)} = (p,q,u,v)$ & $s_{\rm p(s)} = (i,q,u,v)$ \\
$s_{\rm s(p)} = (\hat{n},\hat{n},\hat{n},\hat{n})$ & $s_{\rm s(p)} = 
(\hat{n},\hat{n},\hat{n},\hat{n})$ \\
$s _{\rm u} = (i-p,\hat{n},\hat{n},\hat{n})$ & $s _{\rm u} = 
(\hat{n},\hat{n},\hat{n},\hat{n})$ \\
For $|\phi-\phi_m| < {\pi\over 4}\ (\ge {\pi\over 4})$ & For $|\phi-
\phi_m| < {\pi\over 4}\ (\ge {\pi\over 4})$ \\ \\

$l < 2\sigma$ and $i > 2\sigma$ & $l < 2\sigma$ and $i < 2\sigma$ \\
(depolarisation)  &  (noise)  \\
$s_p,s_s = (\hat{n},\hat{n},\hat{n},\hat{n})$ & $s_p,s_s = 
(i/3,q/3,u/3,v/3)$ \\ 
$s _{\rm u} = (i,q,u,v)$ & $s _{\rm u} = (\hat{n},\hat{n},\hat{n},\hat{n})$ \\
 &  \\
%& \hspace{1.0in} ...(A.2) \\
\end{tabular}
\end{center}

\bigskip

\centerline {\bf References}

\parhang Backer, D. C. 1970{\it a}, {\it Nature}, {\bf 227}, 692. \vskip 
1pt 

\parhang Backer, D. C. 1970{\it b}, {\it Nature}, {\bf 228}, 42. \vskip 1pt 

\parhang Backer, D. C. 1970{\it c}, {\it Nature}, {\bf 228}, 1297. 
\vskip 1pt 

\parhang Backer, D. C. 1971, Ph.D. thesis, Cornell Univ. \vskip 1pt 

\parhang Backer, D. C. 1973, {\it Ap. J.}, {\bf 182}, 245. \vskip 1pt 

\parhang Cole, T. W. 1970, {\it Nature}, {\bf 227}, 788. \vskip 1pt 

\parhang Comella, J. M. 1971, Ph.D. thesis, Cornell Univ. \vskip 1pt

\parhang Cordes, J. M., Rankin, J. M., \& Backer, D. C. 1978, {\it Ap. 
J.}, {\bf 223}, 961. \vskip 1pt

\parhang Deshpande, A. A. 1999, {\it IAU Colloquium \#177} 
Bonn, 149. \vskip 1pt

\parhang Deshpande, A. A. \& Radhakrishnan, V. 1992, {\it JAA}, {\bf 
13}, 151. \vskip 1pt 

\parhang Deshpande, A. A., Ramachandran, R., \& Radhakrishnan, V. 
1999, private communication. \vskip 1pt 

\parhang Deshpande, A. A. \& Rankin, J. M. 1999, {\it Ap. J.}, 
{\bf 524}, 1008. \vskip 1pt

\parhang Drake, F. D., \& Craft, H. D. E. 1968, {\it Nature}, {\bf 220}, 
231. \vskip 1pt 

\parhang Everett, J. E. \& Weisberg, J. M. {\it Ap. J.}, preprint.  

\parhang Ferraro, V. C. A., \& Plumpton, C. 1966, {\it Magneto-Fluid 
Mechanics} (London: Oxford University Press), 23.  \vskip 1pt 

\parhang Gil, J. A. 1981, {\it Acta Physica Polonicae}, {\bf B12}, 1081.  
\vskip 1pt

\parhang Gil, J. A. 1992, {\it A\&A}, {\bf 256}, 495. \vskip 1pt

\parhang Gil, J., Kijak, J., \& Seiradakis, J. H. 1993, {\it A\&A}, {\bf 
272}, 268. \vskip 1pt

\parhang Gil, J. A., Lyne, A. G., Rankin, J. M., Snakowski, J. K., \& 
Stinebring, D. R. 1981, {\it A\&A}, {\bf 255}, 181. \vskip 1pt

\parhang Hankins, T. H., Rankin, J. M., \& Eilek, J. 2000, {\it Ap. J.}, 
in preparation.  \vskip 1pt 

\parhang Izvekova, V. A., Kuz'min, A. D., Malofeev, V. M., \& Shitov, 
Yu. P. 1981, {\it Ap. Space Sci.}, {\bf 78}, 45. \vskip 1pt

\parhang Kramer, M., Wielebinski, R., Jessner, A., Gil, J. A. \& 
Seiradakis, J. A. 1994, {\it A\&ASS}, {\bf 107}, 515. \vskip 1pt

\parhang Komesaroff, M. M. 1970, {\it Nature}, {\bf 225}, 612. \vskip 
1pt 

\parhang Lang, K. R. 1969, {\it Ap. J.}, {\bf 158}, L175. \vskip 1pt 

\parhang Lovelace, R. V., \& Craft, H. D. 1968, {\it Nature}, {\bf
220}, 875. \vskip 1pt

\parhang Malofeev, V. M. 1999, private communication.  \vskip 1pt

\parhang McKinnon, M. M., Stinebring, D. R. 1998, {\it Ap. J.}, 
{\bf 502}, 883. \vskip 1pt

\parhang Mitra, D. \& Deshpande, A. A. 1999, {\it A\&A}, {\bf 346}, 906.
\vskip 1pt

\parhang Narayn, R. \& Vivekanand, M. 1982, {\it A\&A}, {\bf 113}, L3.
\vskip 1pt

\parhang Radhakrishnan, V., \& Cooke, D. J. 1969, {\it Ap. Lett.}, {\bf 
3}, 225. \vskip 1pt

\parhang Rankin, J. M. 1988, {\it Ap. J.} {\bf 325}, 314.  \vskip 1pt

\parhang Rankin, J. M. 1990, {\it Ap. J.} {\bf 352}, 247.  \vskip 1pt

\parhang \_\_\_\_\_\_\_\_ 1993{\it a}, {\it Ap. J.}  {\bf 405}, 285.  
\vskip 1pt

\parhang \_\_\_\_\_\_\_\_ 1993{\it b}, {\it Ap. J. Suppl.}  {\bf 85}, 145.  
\vskip 1pt

\parhang Rankin, J. M., Campbell, D. B., \& Backer, D. C. 1974, {\it 
Ap. J.}, {\bf 188}, 609. \vskip 1pt

\parhang Rankin, J. M., Campbell, D. B., \& Spangler, S. 1975, {\it NAIC 
Report} {\bf \#46}.  \vskip 1pt

\parhang Rankin, J. M. \& Deshpande, A. A. 1999, {\it IAU Colloquium 
\#177} Bonn, 155. \vskip 1pt

\parhang Rankin, J. M. \& Rathnasree, N. 1995, {\it JAA}, {\bf 16}, 327.
\vskip 1pt

\parhang Rankin, J. M. \& Rathnasree, N. 1997, {\it JAA}, {\bf 18}, 91.
\vskip 1pt

\parhang Rankin, J. M., Rathnasree, N., \& and Xilouris, K. 1999, in 
preparation. \vskip 1pt

\parhang Ruderman, M. A. 1972, {\it Ann. Rev. Astr. Ap.}, {\bf
10}, 427. \vskip 1pt

\parhang Ruderman, M. A. \& Sutherland, P. G. 1975, {\it Ap.J.}, {\bf 
196}, 51. \vskip 1pt

\parhang Ruderman, M. A. 1976,  {\it Ap. J.}, {\bf 203}, 206. \vskip 1pt  

\parhang Sieber, W., \& Oster, L. 1975, {\it A\&A} {\bf 38}, 325.  \vskip 
1pt

\parhang Slee, O.B., Mulhall, P.S. 1970, {\it Proc.Astr.Soc.Aust.} {\bf 1}, 322. \vskip 1pt

\parhang Suleymanova, S.A., Izvekova, V.A. 1984, {\it Sov.Astron.} 
{\bf 28}, 53.  \vskip 1pt

\parhang Suleymanova, S.A., Izvekova, V.A, Rankin, J. M. \& 
Rathnasree, N. 1998, {\it JAA}, {\bf 19}, 1. \vskip 1pt

\parhang Sutton, J. M., Staelin, D. H., Price, R. M., \& Weimer, R. 
1970, {\it Ap. J.}, {\bf 159}, L89. \vskip 1pt 

\parhang Taylor, J. H., \& Huguenin, G. R. 1971, {\it Ap. J.}, {\bf 
167}, 273. \vskip 1pt 

\parhang Taylor, J. H., Huguenin, G. R., Hirsch, R. M., \& 
Manchester, R. N. 1971, {\it Ap. Lett.}, {\bf 9}, 205. \vskip 1pt

\parhang Taylor, J. H., Jura, M., \& Huguenin, G. R. 1969, {\it Nature},
{\bf 223}, 797. \vskip 1pt

\parhang Thorsett, S. 1991, {\it Ap. J.}, {\bf 377}, 263.  \vskip 1pt

\parhang Vitkevich, V. V., Alexseev, Yu. I., \& Zhuravlev, Yu. P. 1969, 
{\it Nature}, {\bf 224}, 49. \vskip 1pt 

\parhang Weisberg, J. M., Cordes, J. M., Lundgren, S. C., Dawson, B. R., 
Despotes, J. T., Morgan, J. J., Weitz, K. A., Zink, E. C., \& Backer, 
D. C. 1999, {\it Ap. J. Suppl.}, {\bf 121}, 171.  \vskip 1pt

\end{document}